\documentclass[aps,twocolumn,prx,superscriptaddress,nofootinbib,amsmath,amssymb,floats,floatfix,english]{revtex4-2}
\usepackage{polski}
\usepackage[utf8]{inputenc}
\usepackage[T1]{fontenc}
\usepackage{amsmath}
\usepackage{amssymb}
\usepackage{ifthen}
\usepackage{graphicx}
\usepackage{times}
\usepackage{babel}
\usepackage{color}
\usepackage{bm}
\usepackage[normalem]{ulem}
\usepackage{float}
\usepackage{placeins}
\usepackage{chngcntr}
\usepackage{enumitem}
\usepackage{multirow}
\usepackage{mathtools}

\usepackage{comment}

\newcommand{\ifis}[2]{
\ifthenelse{\equal{#1}{}}{}{#2}
}

\def\be{\begin{equation}}
\def\ee{\end{equation}}
\def\bea{\begin{eqnarray}}
\def\eea{\end{eqnarray}}
\def\bsu{\begin{subequations}}
\def\esu{\end{subequations}}
\def\bi{\begin{itemize}}
\def\ei{\end{itemize}}

\newcommand{\op}[1]{\widehat{#1}}
\newcommand{\dagop}[1]{\widehat{#1}^{\dagger}}

\newcommand{\mc}[1]{{\mathcal{#1}}}

\newcommand{\nonu}{\nonumber}

\newcommand{\ket}[1]{\vert#1\rangle}
\newcommand{\braket}[2]{\langle#1\vert#2\rangle}
\newcommand{\ketbra}[2]{\vert#1\rangle\langle#2\vert}

\newcommand{\average}[1]{\left\langle{#1}\right\rangle_{\rm PP}}
\newcommand{\expect}[1]{\left\langle{#1}\right\rangle}

\newlength{\templength}

\newcommand{\dada}[2]{\frac{\partial{#1}}{\partial{#2}}}
\newcommand{\dd}[2]{\frac{d{#1}}{d{#2}}}
\newcommand{\half}{\ensuremath{\frac{\scriptstyle 1}{\scriptstyle 2}}}

\newenvironment{eq}{\begin{equation}}{\end{equation}}

\newcommand{\eqn}[1]{(\ref{#1})}

\newcommand{\fig}[2]{\hspace*{-0.05\columnwidth}\begin{figure}[htb]\begin{center}\includegraphics[width=\columnwidth]{#1}\end{center}\vspace*{-1em}\caption{#2}\end{figure}\vspace*{1em}}
\newcommand{\figw}[2]{\hspace*{-0.05\columnwidth}\begin{figure*}[htb]\begin{center}\includegraphics[width=\textwidth]{#1}\end{center}\vspace*{-1em}\caption{#2}\end{figure*}\vspace*{1em}}
\newcommand{\figv}[4]{\hspace*{-0.05\columnwidth}\begin{figure}[#3]\begin{center}\includegraphics[width=#4\columnwidth]{#1}\end{center}\vspace*{-1em}\caption{#2}\end{figure}\vspace*{1em}}
\newcommand{\figwv}[4]{\hspace*{-0.05\columnwidth}\begin{figure*}[#3]\begin{center}\includegraphics[width=#4\textwidth]{#1}\end{center}\vspace*{-1em}\caption{#2}\end{figure*}\vspace*{1em}}
\renewcommand{\eq}[2]{\begin{equation}\label{#1}#2\end{equation}}

\newcommand{\eqa}[2]{\begin{eqnarray}\label{#1}#2\end{eqnarray}}
\newcommand{\eqy}[2]{\begin{align}\label{#1}#2\end{align}}

\usepackage{lmodern}

\setlist[itemize]{nosep}
\setlist[enumerate]{nosep,nolistsep}

\newcommand{\REM}[1]{\ifthenelse{0=1}{#1}{}}

\allowdisplaybreaks[3]

\begin{document}

\title{Exact Positive-P Simulation of Open Quantum Spin Systems with Generalized Density Matrix Formalism} 

\author{Alex Ferrier}
\email{(corresponding author) a.ferrier@ucl.ac.uk}
\affiliation{Department of Physics and Astronomy, University College London, Gower Street, London WC1E 6BT, United Kingdom}
\affiliation{Center for Theoretical Physics, Polish Academy of Sciences, Aleja Lotnik\'ow 32/46, 02-668 Warsaw, Poland}

\author{Yi Shi}
\affiliation{Department of Physics and Astronomy, University College London, Gower Street, London WC1E 6BT, United Kingdom}

\author{Marzena H. Szyma\'nska}
\affiliation{Department of Physics and Astronomy, University College London, Gower Street, London WC1E 6BT, United Kingdom}

\author{Piotr Deuar}
\affiliation{Institute of Physics, Polish Academy of Sciences, Aleja Lotnik\'ow 32/46, 02-668 Warsaw, Poland}

\date{\today}
\begin{abstract} 

The positive-P representation for bosons allows scalable exact simulations of many-body open quantum systems, provided dissipation is strong enough to stabilise stochastic trajectories. Designing analogous methods for spin systems, relevant to quantum technologies, remains challenging, as previous spin positive-P formulations can exhibit systematic errors even when simulations are numerically stable. Here, we develop a phase-space representation that maps spin-$\frac{1}{2}$ degrees of freedom onto parameters of a local generalized density matrix (GDM), rather than SU(2) coherent states. In all stable regimes, we observe no deviations from exact solutions, restoring the positive-P property: stable solutions are exact up to stochastic sampling noise. The GDM representation combines seamlessly with bosonic positive-P, enabling highly scalable treatment of coupled spin-boson systems. We map simulation stability versus spin and boson dissipation in the Jaynes-Cummings model and apply the method to a Jaynes-Cummings-Hubbard quasi-Lieb lattice, showing how spin dissipation and driving can destroy or restore excitation incompressibility.

\end{abstract} 

\maketitle

\section*{Introduction}
\label{INTRO}

Spin-$\frac{1}{2}$ systems, commonly referred to as qubits, form the core element for quantum computing and other quantum technologies. Ideally, qubits would possess long \mbox{coherence} times and could be treated as nearly isolated quantum systems. In present-day noisy intermediate-scale quantum (NISQ) devices, however, coupling to the environment remains unavoidable, creating a strong need for computational techniques capable of describing the dissipative dynamics of many-qubit systems. Candidate platforms for quantum computation and simulation, \mbox{including} cavity \cite{Raimond01r,Walther06r,Reiserer15r,Lepert2011} and superconducting circuit QED \cite{Houck12,Schmidt13,Fitzpatrick17,Fink17,RevModPhys.93.025005}, can be viewed as open quantum systems consisting of arrays of coupled spins and bosons \cite{Noh16}.

With direct solution of the quantum dynamics {limited} by the exponential growth of the Hilbert space with system size, recent efforts to meet the demand of \mbox{experimental} progress have seen an explosion in {different} approaches for realising efficient and {scalable} {numerical} simulations of open quantum spin and \mbox{spin-boson} {arrays.}  One direction has been to extend methods successful for closed spin systems to include dissipation, as achieved \mbox{recently} with tensor-network \cite{PhysRevX.11.021035,dunham2025,Hryniuk2024,hryniuk2026} and neural-quantum-state \cite{PhysRevLett.122.250501,PhysRevLett.122.250502,PhysRevLett.122.250503} approaches.  \mbox{Tensor} \mbox{networks} can be exceptionally powerful, but their {applicability} is often constrained by \mbox{geometry} or \mbox{dimensionality,} or limited to short-range couplings.  \mbox{Neural} quantum states offer a complementary route, however their {scalability} for open-system dynamics \mbox{remains} limited.

\figwv{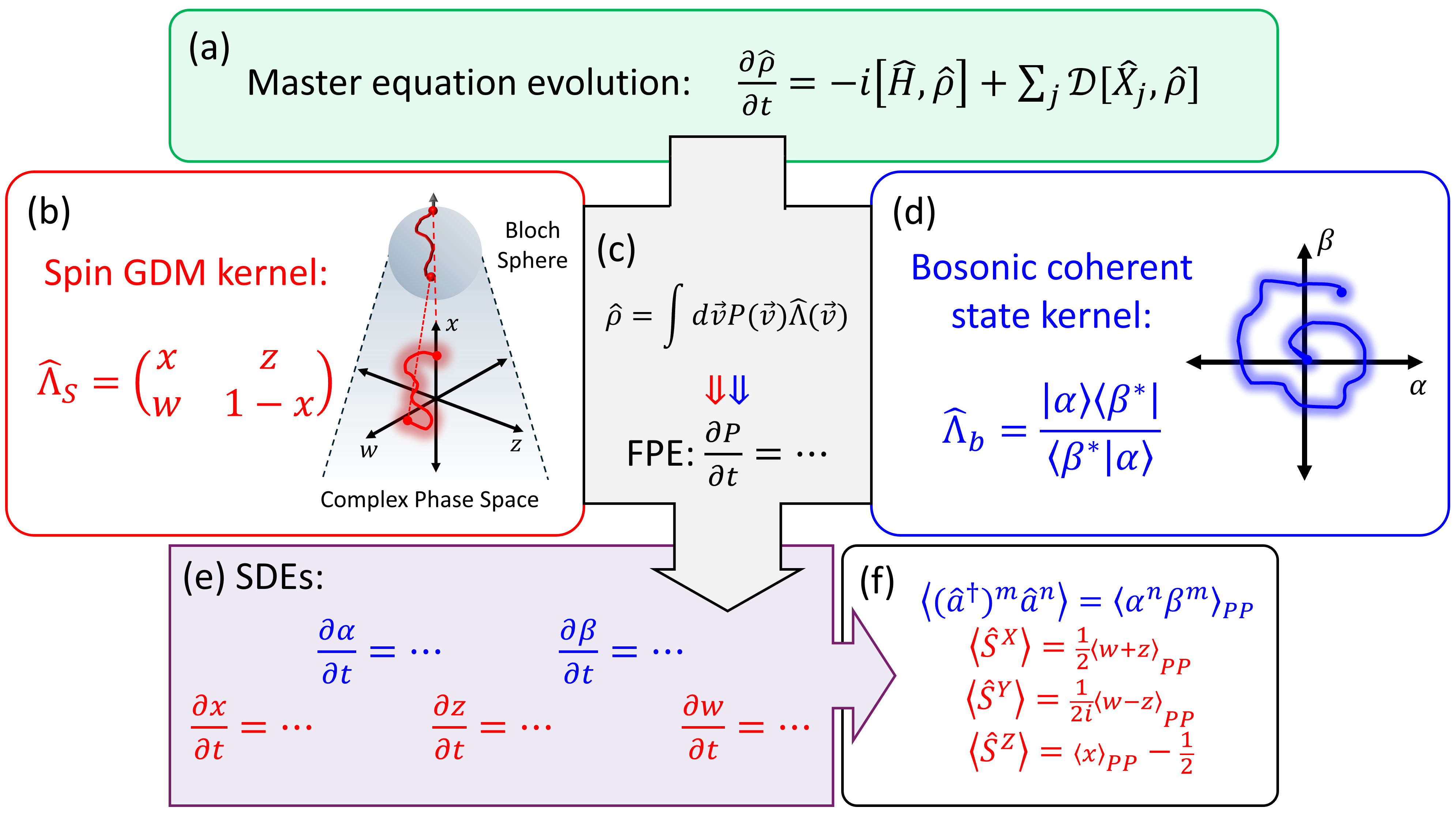}{
{\bf Method Diagram.} a) A Markovian open quantum system with density matrix $\op{\rho}$ evolves according to a Lindblad master equation. b) Spin-$\frac{1}{2}$ states are mapped to a three-variable complex phase space using the generalized density matrix (GDM) kernel $\op{\Lambda}_S$. c) Kernel identities transform the master equation into a Fokker--Planck equation (FPE) for the positive-P distribution $P$. d) For coupled spin-boson systems, the mapping is combined with the bosonic positive-P coherent state kernel $\op{\Lambda}_b$, with $\op{\Lambda}=\op{\Lambda}_S\otimes\op{\Lambda}_b$ and $\vec{v}={x,z,w,\alpha,\beta}$. e) The FPE is converted into stochastic differential equations (SDEs) generating trajectories that sample $P$. f) Quantum observables are obtained from ensemble averages over these trajectories.  \label{GDM_leadfig}}{htb}{0.95}

Alternatively, \emph{stochastic phase-space methods}, including the truncated Wigner \cite{Norrie05,Sinatra02,Martin10b,Steel98,Deuar09b,Hoffmann08,Carusotto05,Wouters09,Dagvadorj15,Dominici15,Donati16,Caputo17,Comaron18,Ballarini20} and \mbox{positive-P} \cite{Drummond80,Carter87,Lewis-Swan14,Deuar07,Kheruntsyan12,Gilchrist97,PRXQuantum.2.010319} representations, map the density matrix evolution onto stochastic \mbox{trajectories} in a complex phase space. These methods were originally developed for bosonic \mbox{systems,} with the phase spaces constructed from coherent states. By exchanging the computational \mbox{complexity} of \mbox{handling} \mbox{exponentially} large density matrices for \mbox{averaging} over stochastic trajectories, they can achieve exceptional \mbox{scalability:} full quantum dynamics of systems containing up to millions of modes have been demonstrated \cite{Deuar07,Kheruntsyan12}, without requiring a particular geometry, dimensionality, interaction range, or spatial homogeneity. For the positive-P representation, the mapping is exact. Its principal limitation is instead \mbox{dynamical:} stochastic noise can strongly amplify itself such that trajectories can escape to infinity, \mbox{restricting} accessible evolution times.  When trajectories remain numerically stable, \mbox{enabled} by sufficiently strong dissipation, \mbox{positive P} should converge to the exact quantum solution without \mbox{systematic} errors or \mbox{approximations} \cite{PRXQuantum.2.010319}.

The versatility of stochastic phase-space methods has motivated considerable interest in extending them from bosonic to spin systems, including discrete truncated Wigner \cite{PhysRevX.5.011022,Zhu19,PhysRevA.105.013716,Bond24}, continuous-variable Wigner methods for spins \cite{Hartmann2026arxiv,Hosseinabadi25}, and stochastic approaches based on other theoretical constructions \cite{PhysRevB.104.024408,PhysRevResearch.5.043288,PRXQuantum.4.030304,PhysRevX.15.021073}. Achieving exact and scalable positive-P simulations for spins has proved more difficult. Previous formulations have either focused on closed spin chains \cite{Barry08,Ng11,Ng13}, with no dissipation to \mbox{stabilise} the trajectories, or can develop uncontrolled systematic errors in observables, so-called boundary-term errors, \mbox{before} any stochastic noise instability becomes apparent \cite{Mandt15,Chuchurka24b,Leonau25arxiv}. Thus, unlike the established bosonic \mbox{positive-P} representation, numerical stability has not guaranteed exact \mbox{results} for spins.

Here, we first identify how these difficulties arise in the previous SU(2) coherent-state-based positive-P mappings \cite{Ng13,Mandt15}. Guided by this analysis, we \mbox{develop} a phase-space representation for spin-$\frac{1}{2}$ systems based \mbox{instead} on a parametrisation of a local generalized density matrix (GDM). The construction does not rely on spin coherent states, but still integrates naturally with the conventional coherent-state-based bosonic \mbox{positive-P} representation, allowing coupled spin-boson systems to be treated within a single framework. Because the \mbox{number} of stochastic complex variables grows only \mbox{linearly} with system size, the method retains the high scalability characteristic of phase-space approaches. The construction is summarised in Fig.~\ref{GDM_leadfig}.

We then show that the GDM positive-P representation gives exact results whenever the stochastic \mbox{trajectories} remain numerically stable, with no statistically significant deviations from exact quantum solutions in the \mbox{parameter} regimes tested.  This provides a scalable route to quantitatively reliable simulations of dissipative spin-boson dynamics. Finally, we use the method to address a many-body physics problem, investigating how the incompressibility of combined spin and boson excitations previously found in a Jaynes-Cummings-Hubbard quasi-Lieb lattice \cite{Biondi15} is modified, destroyed, or restored by spin driving and dissipation.
 
\section*{Results}
\subsection{Model}
\label{model}
We explore the application of the method to models of coupled spins and bosons with drive and \mbox{dissipation.}  As an archetypal example of this class of systems, we consider a driven-dissipative Jaynes-Cummings (DDJC) model (see Fig.~\ref{DDJC_sys}a) with the following local \mbox{Hamiltonian}:
\eqa{H_JC}{
\op{H}^{JC}_j &= \omega_C\dagop{a}_j\op{a}_j +\omega_0\op{S}_j^Z +g\left(\op{a}_j\op{S}_j^+ +\dagop{a}_j\op{S}_j^-\right) \nonu \\ &+h\op{S}_j^X +F\left(\op{a}_j+\dagop{a}_j\right)\, ,
}
where each site $j$ has a local bosonic mode with annihilation operator $\op{a}_j$, coupled to a qubit with spin projection operators 
\eq{SpinMat}{
\op{S}_j^X = \frac{1}{2}\!\begin{pmatrix} 0 & 1 \\ 1 & 0 \end{pmatrix}, \, \op{S}_j^Y = \frac{1}{2}\!\begin{pmatrix} 0 & -i \\ i & 0 \end{pmatrix}, \, \op{S}_j^Z = \frac{1}{2}\!\begin{pmatrix} 1 & 0 \\ 0 & -1 \end{pmatrix}, \nonu
}
written in the basis $\ket{\!\!\downarrow\,}_j=\begin{psmallmatrix}0 \\ 1\end{psmallmatrix}$, $\ket{\!\!\uparrow\,}_j=\begin{psmallmatrix}1 \\ 0\end{psmallmatrix}$, along with spin raising/lowering operators $\op{S}_j^\pm = \op{S}_j^X \pm i\op{S}_j^Y$.  Here, $\omega_C$ and $\omega_0$ are the energy of the bosons and spins, \mbox{respectively,} written relative to the driving frequency. $g$ is the Jaynes-Cummings (JC) interaction, and the system is driven either via a coherent drive $F$ on the bosons or a transverse field $h$, which drives oscillations of the spin directly.  Multiple sites are coupled by a hopping $J_{j,j'}$ \mbox{between} the bosonic modes of connected sites $j,j'$, leading to a driven-dissipative Jaynes-Cummings-Hubbard (DDJCH) model (see Fig.~\ref{DDJC_sys}b) with Hamiltonian
\eq{H_JCH}{
\op{H} = \sum_j H^{JC}_j \,-\!\!\!\!\!\!\sum_{\mathrm{connections}\, j,j'} \! \left[J_{j,j'}\dagop{a}_j\op{a}_{j'} +J_{j,j'}^*\dagop{a}_{j'}\op{a}_j \right]\, .
}
We consider three basic forms of local dissipation: bosonic dissipation $\gamma$, spin decay $\Gamma_\downarrow$, and spin dephasing $\Gamma_\phi$.  These appear in the evolution of the density matrix $\op{\rho}$, which describes the state of the system, via the master equation as 
\eqa{ME}{
&\dd{\op{\rho}}{t} = -i[\op{H}, \op{\rho}\,]  \\ &+\sum_j \left(\frac{\gamma}{2}\mc{D}[\op{a}_j,\op{\rho}\,] +\frac{\Gamma_\phi}{2}\mc{D}[\op{S}^Z_j,\op{\rho}\,] +\frac{\Gamma_\downarrow}{2}\mc{D}[\op{S}^-_j,\op{\rho}\,]\right)\, , \nonu 
}
where $\mc{D}[\op{X}, \op{\rho}\,] \equiv 2\op{X}\op{\rho}\dagop{X} - \dagop{X}\op{X}\op{\rho} -\op{\rho}\dagop{X}\op{X}$ is shorthand for the Lindbladian dissipator generated by $\op{X}$.  

In principle, the GDM method is also applicable to pure spin lattices with sufficient dissipation. However, in this work, we focus on coupled spin-boson models, as this allows us to address one of the main applications and motivations for a tractable continuous-variable representation: handling both spin and bosonic degrees of freedom seamlessly within the same numerical framework.

\figv{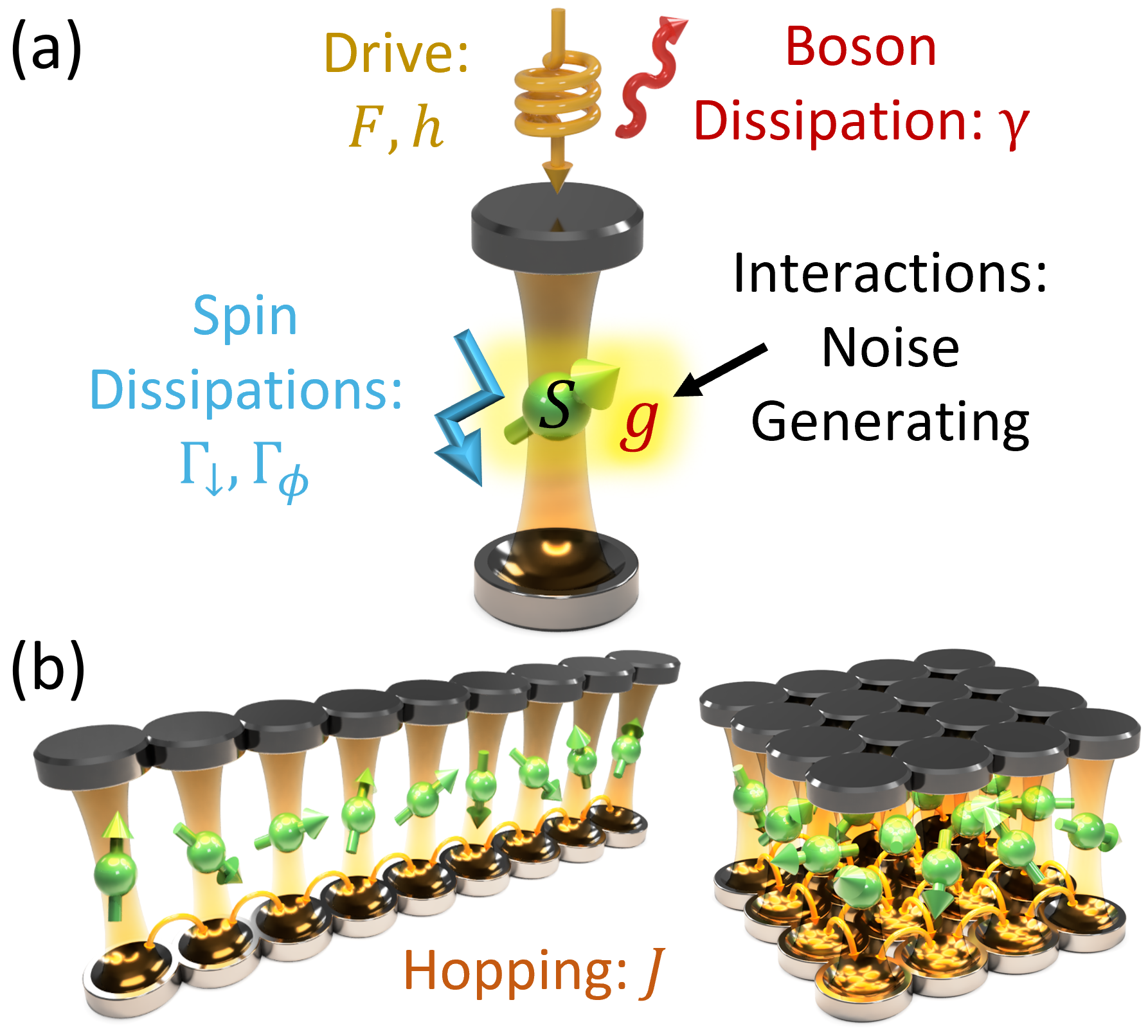}{{\bf Illustration of the model. } a) Diagram of a single site driven-dissipative Jaynes-Cummings (DDJC) system, with driving terms ($F$, $h$) and dissipation terms ($\gamma$, $\Gamma_\downarrow$, $\Gamma_\phi$) for bosons and spins respectively, and spin-boson coupling $g$, which is the source of noise in our positive-P representation.  b) DDJC chains and lattices with bosonic hopping $J$.  \label{DDJC_sys}}{t}{1}

\vspace{-5mm}
\subsection{Generalizing Positive P to spins}
\label{PP}

At the most general level, the positive-P representation can be viewed as a mapping between the density matrix $\op{\rho}$ and a probability distribution $P(\vec{v})$ over a set of complex phase-space variables $\vec{v}$, using a kernel operator $\op{\Lambda}(\vec{v})$:
\eqa{PPdef}{
\op{\rho} = \int d\vec{v}\  P(\vec{v})\op{\Lambda}(\vec{v}) \, , \\
\op{\Lambda}(\vec{v}) = \bigotimes_m \op{\Lambda}_m(\vec{v}_m) \, , \nonu
}
with $\vec{v}$ being the set of phase space variables for the \mbox{entire} \mbox{system,} and $\vec{v}_m$ the subset of variables corresponding to subsystem $m$.  The $\op{\Lambda}_m(\vec{v}_m)$ is the local kernel operator for subsystem $m$.  From the perspective of numerical methods, the utility of this formalism lies in the fact that the properties of the kernel operator can be used to transform the quantum master equation governing the evolution of $\op{\rho}$ into a Fokker--Planck equation (FPE) for the distribution $P(\vec{v})$. In particular, for Hamiltonians containing interactions up to two-body order, and for the dissipative processes considered here, the positive-$P$ representation yields an exact FPE involving derivatives of at most second order. This is in contrast to, for example, the Wigner representation, for which nonlinear interactions generally give rise to higher-order derivative terms that must be neglected to obtain the truncated-Wigner approximation. The resulting FPE can then be mapped onto stochastic differential equations describing trajectories in phase space that sample the distribution.

Here, we investigate systems where both bosonic and spin-\half\ subsystems are coupled.  For the bosonic modes, we can use the usual well-established \mbox{kernel} operator for bosonic positive P \cite{Drummond80,PRXQuantum.2.010319} on a phase space of complex coherent state amplitudes $(\alpha,\beta)$:
\eq{bosonLambda}{
\op{\Lambda}_b = \frac{\ketbra{\alpha}{\beta^*}}{\braket{\beta^*}{\alpha}}\, ,  \ket{\alpha} = e^{-\frac{1}{2}\left|\alpha\right|^2}\exp\!\left[{\alpha\dagop{a}}\right]\ket{0} \, , 
}
with $\ket{0}$ denoting the bosonic vacuum state. One may also map the spins onto bosons using the Schwinger-boson approach, allowing them to be treated with the same kernel operator \cite{Ng11}. However, our initial investigations of this approach revealed errors in the treatment of spin dissipation, as well as difficulties in consistently incorporating all relevant initial conditions. These issues ultimately indicated that the approach would not provide a sufficiently general or robust framework for our purposes (for details, see section~\ref{Schwing}). The primary aim of this work is therefore to construct an appropriate and well-behaved local kernel operator for the spin degrees of freedom, such that both spin and bosonic degrees of freedom can be treated numerically within a compatible formalism.

\subsubsection{SU(2) coherent state kernel}

A first approach to achieving this, explored in previous work \cite{Ng13,Mandt15}, is to emulate the bosonic kernel \eqn{bosonLambda} using the spin analogue of bosonic coherent states. These so-called SU(2) coherent states can be defined in direct analogy with their bosonic counterparts, using the exponential of the spin-raising operator $\op{S}^+$ parametrized by a complex number $z$:
\eq{SU2c}{
\ket{z} = \frac{\exp\!\left[z\op{S}^+\right]}{1+\left|z\right|^2}\ket{\!\downarrow\,}\, .
}
The spin kernel operator is then constructed as in eq.~\eqref{bosonLambda}, but from a pair of these SU(2) coherent states $(z,w)$:
\eq{SU2k}{
\op{\Lambda}_S = \frac{\ketbra{z}{w^*}}{\braket{w^*}{z}} = \frac{\exp\!\left[z\op{S}^+\right]\ketbra{\!\downarrow\,}{\,\downarrow\!}\exp\!\left[w\op{S}^-\right]}{1+zw}\, .
}
It should be noted that there is a complication that arises if we want to use the phase space defined by these SU(2) coherent states for numerical simulations, in that if for example the spin down state ($\op{S}^Z\ket{\!\!\downarrow}=-\frac{1}{2}\ket{\!\!\downarrow}$) is represented at $z=0$, as in the construction in \eqn{SU2c}, the opposite state on the Bloch sphere ($\op{S}^Z\ket{\!\!\uparrow}=\frac{1}{2}\ket{\!\!\uparrow}$) must be represented by $z\to\infty$.  A way around this in simulations is to switch to the opposite mapping with spin up at zero dynamically when trajectories cross the equator of the Bloch sphere \cite{Ng13}.  Further details of the stochastic equations that result from using this kernel operator for spins coupled to bosons can be found in section~\ref{SESU2}.  

Although theoretically this allows us to extend the \mbox{positive-P} representation to spins in a formally scalable manner (the number of complex variables per sample for a system with $N$ bosonic modes and $N$ spins is only $4N$), in actual simulations we find that it can encounter systematic errors in a way that undermines the intended advantages of the method. While these \mbox{issues} were not encountered for closed spin chains \cite{Barry08,Ng13}, both our preliminary investigations (see \mbox{section~\ref{SESU2}}) and previous work \cite{Mandt15} on dissipative \mbox{spin-boson} models have shown that, for some parameter regimes, simulations can remain numerically stable while exhibiting systematic errors relative to the exact operator solution of the master equation. We reiterate that a key benefit of the \mbox{positive-P} method for exact and scalable simulations of many-body systems in bosonic or closed spin systems has been that failure of the method is clearly signalled by the stochastic trajectories becoming unstable, or at least by the appearance of indicators of the onset of such instability, such as spikes \cite{Gilchrist97,Deuar02,Yi_arXiv}. If systematic errors can occur without these signals, then even if the magnitude of these errors is small, we can no longer be confident that the simulations continue to yield accurate \mbox{values} for \mbox{quantum} observables when scaled up to system sizes for which benchmarking against other exact methods \mbox{becomes impossible.}

To see how these issues arise, we look in more detail at the SU(2) coherent state kernel operator.  For this, and for the developments in subsequent sections, it is helpful to write the kernel operator as a $2\times2$ matrix in the space of spin vectors:
\eq{SU2kernM}{
\op{\Lambda}_S = \frac{1}{1+zw}\begin{pmatrix} zw & z \\ w & 1\end{pmatrix}\, .
}
It can be seen that a singularity occurs in the kernel operator whenever $zw=-1$, where the trace-normalising factor in the denominator vanishes. Since the density matrix itself is finite, this means, by the definition \eqn{PPdef}, that a correct phase-space distribution $P(z,w)$ must vanish at these points. Consequently, the probability density is also very small in their vicinity. While in the idealised limit of infinitely many stochastic samples the mapping \eqn{PPdef} still holds and the correct results would be recovered, in practice a finite ensemble may contain very few, or even no, trajectories that pass sufficiently close to these regions. Such rare trajectories can nevertheless carry a disproportionately large weight in the stochastic averages because of the singular behaviour of the kernel, so that failing to sample them adequately leads to a systematic bias in the simulation results.
Moreover, in a case such as \eqn{SU2kernM}, where the singular points occur at finite values of the phase-space variables, rather than only at infinity as for the bosonic kernel \eqn{bosonLambda}, their vicinity can be readily reached or even inaccurately crossed due to the dynamical noise in the evolution equations. Numerically, this noise necessarily produces finite random kicks of size $\mc{O}(\sqrt{\Delta t})$, which decrease only slowly with the time step $\Delta t$.

\subsubsection{Generalized density matrix kernel}

In order to restore the unbiased convergence of the \mbox{positive-P} method to the exact solution for finite stochastic samples in our treatment of the spin degrees of freedom, we must modify the spin kernel operator so that it no longer contains such singularities. The construction shown in eq.~\eqn{SU2kernM} illuminates a way forward in this regard: since the spin kernel can be expressed as a $2\times2$ matrix, it provides a convenient route to considering more general spin kernels that cannot be written as outer products of basis states, such as the SU(2) coherent states, as has been done almost exclusively to date. We therefore modify the form of \eqn{SU2kernM} so that its trace is unity by construction, without requiring division by a normalising factor that could potentially lead to singularities. To achieve this, we introduce a third complex phase-space variable $x$ (as \mbox{illustrated} in Fig.~\ref{GDM_leadfig}(b)), which appears in both diagonal elements, leading to a new three-variable spin kernel operator: \pagebreak
\eq{xkern}{
\op{\Lambda}_S = \begin{pmatrix} x & z \\ w & 1-x\end{pmatrix}\, .
}
An additional benefit of this kernel operator compared to the SU(2) kernel is that we can now represent all points on the Bloch sphere with finite values in our phase space: the spin down state is at $x=z=w=0$, while spin up is at $x = 1, z=w=0$.
In fact, \eqn{xkern} has the form of a completely general local density matrix, except that the complex $x$ and independence of $z$ and $w$ may allow it to be non-Hermitian, so any valid spin state, pure or mixed, can be represented by finite fixed values of $x,z,w$.  We therefore refer to it as a generalized \mbox{density} matrix (GDM) kernel.  The derivation of \mbox{stochastic} \mbox{differential} equations using this kernel operator, also including the definition of the diffusion gauge $G$, are given in the \mbox{methods} section~\ref{sec:GDMSDE}.

\subsubsection{Observables}
\label{ObsSec}

As usual with the positive-P method, quantum observables ($\expect{...}$) are calculated as appropriate averages over the stochastic trajectories ($\average{...}$).  Bosonic properties naturally continue to be calculated in the same way as is known for the bosonic positive-P kernel e.g.~ 
\eqy{bosObs}{
n_j&=\expect{\dagop{a}_j\op{a}_j}=\average{\alpha_j\beta_j}\, , \nonu\\
g^{(2)}_j(0) &= \expect{\dagop{a}_j\dagop{a}_j\op{a}_j\op{a}_j}/n_j^2=\average{\alpha^2\beta^2}/n_j^2\,;
}
expectations of spin operators can be \mbox{calculated} from the new phase space as follows: 
\eqy{spinObs}{
\expect{\op{S}^Z_j} =& \average{x_j}-\frac{1}{2}\, , \nonu\\
\expect{\op{S}^-_j} = \average{z_j}\, , &\quad 
\expect{\op{S}^+_j} = \average{w_j}\, .
}
We note that while trajectory-wise $z_j$ and $w_j$ are independent, from the symmetries of the stochastic equations \eqn{Ito_full} we should still have $\average{z_j} = \average{w_j}^*$ when averaging over sufficiently many samples drawn from an initially Hermitian distribution, preserving the Hermitian conjugate relation between the operators $\expect{\op{S}^-_j}=\expect{\op{S}^+_j}^*$.  

Higher-order correlations also take simple forms. In particular, since the positive-P representation naturally yields normally ordered expectation values, rather than the symmetrically ordered moments associated with the Wigner representation, higher-order correlation functions can be evaluated straightforwardly. For \mbox{example:}
\eqa{SzSz}{\expect{\op{S}^Z_j\op{S}^Z_k} &=& \average{\left(x_j-\frac{1}{2}\right)\left(x_k-\frac{1}{2}\right)}\\
\expect{\op{S}^+_j\op{S}^-_k} &=& \average{w_jz_k}.}
These and others are obtained by evaluating the trace of the product of the corresponding observable with \eqn{xkern}.

\subsubsection{Dynamics}
\label{DynSec}

The stochastic equations for all phase-space variables resulting from the Hamiltonian and dissipative dynamics of the model are described in the Methods section~\ref{sec:GDMSDE}. The final equations that we simulate are given in \eqn{Ito_full}. In these equations, a diffusion stochastic gauge \cite{Deuar02,Deuar06b}, $G$, has been introduced to control the balance of noise associated with the spin-boson interactions (large $G$ corresponds to more bosonic noise and less spin noise, while small $G$ \mbox{corresponds} to relatively more spin noise). It does so without affecting the asymptotic means of \mbox{observables} as the number of trajectories increases. The gauge can, however, affect the signal-to-noise ratio and the robustness of the simulation against noise amplification.  We compare the simulated dynamics obtained using the two different spin kernels in Fig.~\ref{xKernComp}.  Remarkably, the GDM kernel gives exact results in a case where the SU(2) coherent state kernel produces strong systematic errors.  We will find in the following section that the GDM approach in fact \emph{always} gives exact results for \mbox{quantum} observables in regimes where the simulations do not \mbox{suffer} a noise catastrophe.  

\fig{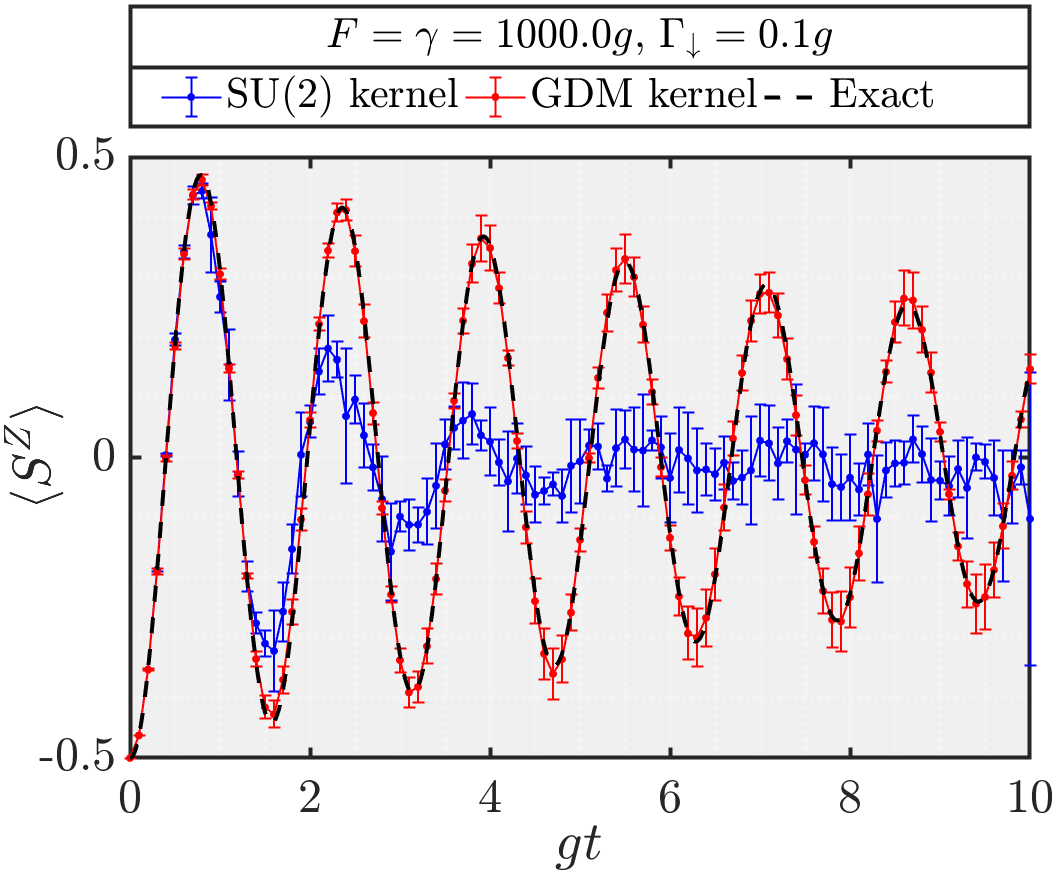}{{\bf Comparison of the SU(2) and GDM kernel.}  Comparison of $\expect{\op{S}^Z}$ between simulations made using the SU(2) kernel ((\ref{SU2kernM}) - blue line) and the GDM kernel ((\ref{xkern}) - red line) for a single site with $F = \gamma = 1000g$, $\Gamma_\downarrow = 0.1g$, where the SU(2) kernel results show systematic errors.  Other parameters $h$, $\omega_C$, $\omega_0$, $\Gamma_{\phi}$ are 0.  The GDM kernel matches the exact solution (black dashed line) for as long as it is stable.  The GDM kernel simulation here uses a diffusion gauge $G = 10$.  \label{xKernComp}}

\subsection{Regimes of stability and accuracy}
\label{Stab}
 
Analogously to the purely bosonic case, we expect that the stability of the positive-P method depends on the interplay between the Jaynes-Cummings interaction $g$, which generates noise in the equations \eqn{Ito_full}, and the various forms of dissipation which should act to suppress the noise amplification instabilities.  Following a similar analysis to \cite{PRXQuantum.2.010319} and \cite{Yi_arXiv}, we investigate the stability of the GDM positive-P method across orders of magnitude in the relative strengths of the dissipations compared to the interactions: $\gamma/g$ and $\Gamma_\downarrow/g$.  We first simulate the case of a single site Jaynes-Cummings model (one spin coupled to one bosonic mode), as this allows us to easily benchmark our method against direct operator solutions of the master equation, to show that the stable cases are indeed exact to the underlying quantum mechanics.

In this context, we note that previous studies have confirmed that, for local interactions and losses, the stability and effective maximum simulation time of larger systems are determined almost entirely by the single-site stability conditions \cite{Deuar06a,Deuar06b,ppdd}.  In what follows, we compare the \mbox{observables} $n$ and $\expect{\op{S}^Z}$ with the exact results. The drive strength is fixed at $F=\gamma$ throughout these investigations, so as to bound the maximum bosonic occupation introduced into the system. This \mbox{allows} us to avoid \mbox{instabilities} arising primarily from \mbox{excessively} strong driving, as previously observed in bosonic \mbox{models} \cite{PRXQuantum.2.010319}, and \mbox{instead} focus on those resulting from the \mbox{balance} of interactions and dissipation. The remaining parameters are $h=\Gamma_\phi=\omega_0=\omega_C=0$. We run 1000 trajectories for each case, up to a time of at least $gt=100$, or $gt=1000$ in the few cases where a longer evolution is required to reach a steady state.

\vspace{-5mm}
\figv{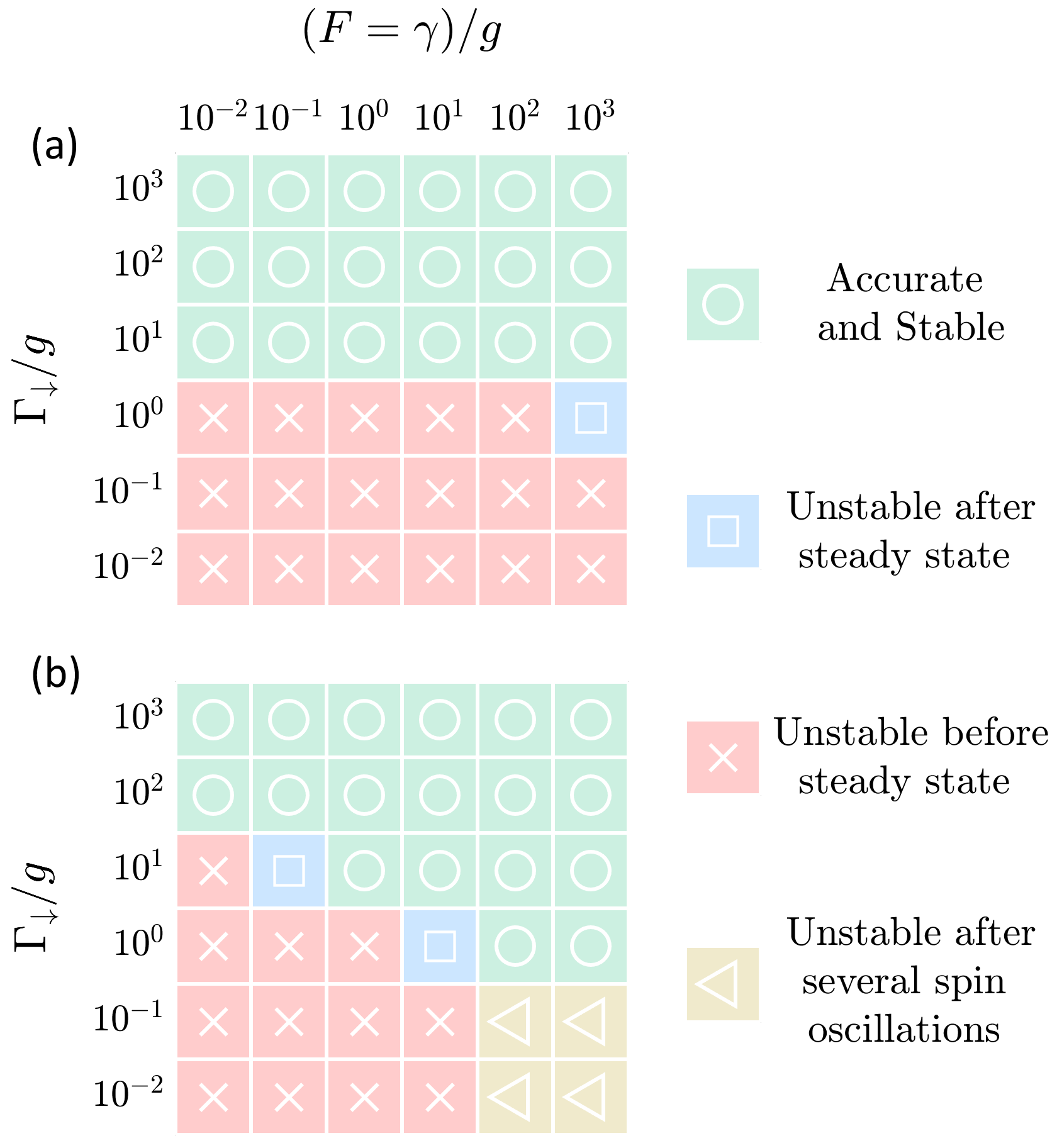}{{\bf Stability diagrams for the GDM kernel.}  \mbox{Stability} of simulations across orders of magnitude in $\gamma$ and $\Gamma_\downarrow$ for a) $G=1$ (no diffusion gauge) and b) $G=10$. Red squares indicate simulations that are initially accurate within statistical uncertainty but become unstable before reaching the steady state; green, simulations that remain accurate and \mbox{stable} throughout; blue, simulations that reach the steady state accurately but become unstable at later times; and \mbox{yellow,} simulations that remain stable and accurate for several spin oscillations without reaching the steady state.  Unlike in earlier studies using other kernels \cite{Mandt15,Chuchurka24b,Leonau25arxiv}, no statistically significant deviations from the exact solution were observed for any set of parameters.  \label{StabGDM}}{t}{1}

\figw{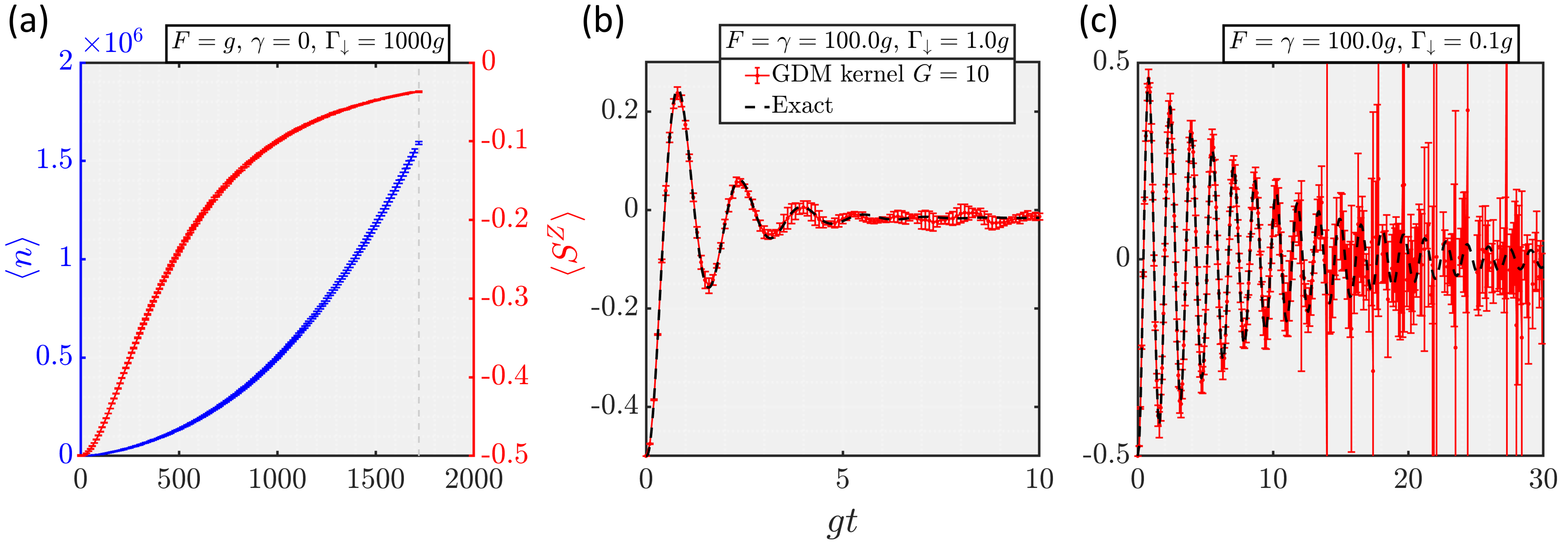}{{\bf Examples of stability regimes for GDM kernel simulations.} a) Time evolution of boson occupation $n$ and $\expect{\op{S}^Z}$ for $\gamma=0$, $\Gamma_\downarrow=1000g$, and $F=g$ (all other parameters zero). \mbox{Trajectories} become unstable after $gt=1700$, before $n$ reaches a steady state, as in the red squares in Fig.~\ref{StabGDM}. b) $\expect{\op{S}^Z}$ for $F=\gamma=100g$, $\Gamma_\downarrow=g$, and diffusion gauge $G=10$. Positive-P simulations (red line) remain stable and follow the exact solution (black dashed line) throughout; green square in Fig.~\ref{StabGDM}. c) $\expect{\op{S}^Z}$ for $F=\gamma=100g$, $\Gamma_\downarrow=0.1g$, and $G=10$; yellow square in Fig.~\ref{StabGDM}(b). Simulations become unstable before reaching steady state, but remain stable and accurate for several oscillations.  \label{StabEx}}

\vspace{-5mm}
To begin with, we test the stability for the case of the ``default'' diffusion gauge $G = 1$.  A summary of these results are shown in Fig.~\ref{StabGDM}(a).  While we have noticed that the presence of at least some dissipation on both the spins and the bosons is required for the simulations to remain stable indefinitely (see Fig.~\ref{StabEx}(a) for an example of the issues that arise for $\gamma = 0$), for $G =1$ the strength of the spin decay $\Gamma_\downarrow$ plays a much stronger role than the bosonic dissipation $\gamma$ in determining which parameter regimes are stable.  This can be understood in the context of the stochastic equations \eqn{Ito_full} in which only the phase space variables corresponding to the spin degrees of freedom $(z, w, x)$ multiply the noise terms and therefore can directly feedback their amplitude into the variance of the noise terms, possibly promoting the noise amplification instability.  Broadly, we find that for $G = 1$ the simulations are stable indefinitely when $\Gamma_\downarrow>g$.  

We now explore how diffusion gauges (see Methods \mbox{section}~\ref{sec:DiffuG} for details) can be used to expand the \mbox{parameter} regimes in which the method remains \mbox{useful.} We repeat the above analysis using a different \mbox{constant} diffusion gauge, $G=10$, with the corresponding \mbox{stability} diagram shown in Fig.~\ref{StabGDM}(b). Having $G>1$ effectively shifts some of the noise from the spin variables to the bosonic ones. As seen in Fig.~\ref{StabGDM}, this alters the \mbox{stable} \mbox{region} relative to the $G=1$ case, stabilising some borderline cases with large $\gamma$ and $\Gamma_\downarrow\sim g$ at the expense of destabilising some cases with small $\gamma$ that were previously \mbox{stable}. Fig.~\ref{StabEx}(b) shows an example of a case stabilised in this way.  An additional notable feature is the set of cases marked by yellow squares in Fig.~\ref{StabGDM}(b). Due to the small value of $\Gamma_\downarrow/g$, the spin \mbox{undergoes} many oscillations before reaching a steady state. Although the GDM \mbox{positive-P} simulations do not remain stable long enough to resolve the steady state well, they persist for several oscillations and can therefore still be useful for studying the transient dynamics. An example is shown in Fig.~\ref{StabEx}(c).
Overall, this study shows that by \mbox{adjusting} the diffusion gauge \mbox{appropriately,} the method can be made useful over a substantially broader range of parameters than with a single fixed choice of $G$. More sophisticated optimisation of the gauge may extend this range further; indeed, recent work for bosonic systems has explored the use of neural networks to optimise stochastic gauges \cite{cao2026neuralgaugeprepresentationopen}. \mbox{Crucially,} in contrast to the \mbox{SU(2) kernel,} simulations using the GDM kernel remain accurate to within statistical uncertainty whenever they are \mbox{stable}. Consequently, when scaling to larger systems beyond the reach of other exact methods, we can retain confidence in the calculated observables.

\subsection{Scaling to extended lattices}
\label{Latt}

Given that one of the major advantages of stochastic phase-space methods is their ease of scaling to large system sizes, we now turn to demonstrating the use of our method for studying many-site DDJCH lattices. As an example, we consider the \mbox{quasi-Lieb lattice} structure explored in a similar form by Biondi {\it et al.} \cite{Biondi15}, consisting of a 1D DDJCH chain in which only the odd ($A$) sites are coupled to a spin ($Q$), while the even ($B$) sites consist only of a linear bosonic mode with drive and dissipation (see Fig.~\ref{ABQLieb}). Each pair of sites together with the coupled spin is roughly equivalent to the unit cell of a Lieb lattice, with the spin playing the role of the third site and the JC coupling that of the hopping to it. Compared with the work of Biondi {\it et al.}, we additionally include a drive $h$ and dissipation $\Gamma_\downarrow$ on the spins. The latter is required for the stability of our method, while the inclusion of both terms also provides a novel extension in which to investigate how their results are modified by direct driving and dissipation of the spins.

\fig{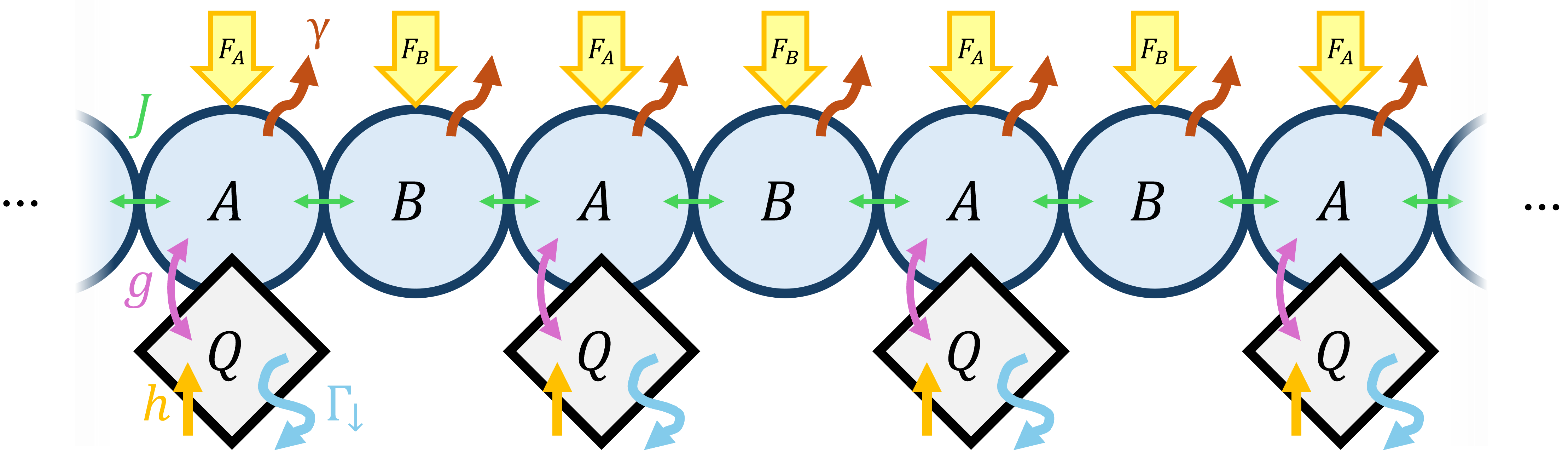}{{\bf DDJCH model in a pseudo-Lieb lattice structure.} Diagram of a 1D DDJCH chain in which only the odd ($A$) sites are coupled to a spin ($Q$). This is analogous to a Lieb lattice with unit cells consisting of two bosonic modes, $A$ and $B$, and a spin qubit $Q$ coupled to the $A$ site, as explored in \cite{Biondi15} but with additional drive and dissipation on the spins.  \label{ABQLieb}}

For this, we simulate a chain of 20 unit cells with periodic boundary conditions. At this system size, a direct treatment of the full density matrix is generally intractable owing to the exponential growth of the many-body Hilbert space and the additional local truncation required for the bosonic modes. Tensor-network methods may access comparable system sizes in one dimension, although periodic boundary conditions and driven-dissipative dynamics can substantially increase their computational cost. The translational invariance of the system allows us to further improve the stochastic sampling by spatially averaging observables over the unit cells in addition to averaging over trajectories.
Here, we investigate several observables beyond those defined in Sec.~\ref{ObsSec}.  First, we study the main signatures of ``incompressibility'' identified in Ref.~\cite{Biondi15}: the mean \mbox{number} of excitations per mode, $\nu$, including both bosonic and spin excitations, and the number fluctuations, $K$. In the Positive-P representation, these are \mbox{defined} and \mbox{evaluated} as
\eqa{LiebNume}{
\nu =& \frac{1}{3}\expect{\dagop{a}_{A}\op{a}_{A}+\dagop{a}_{B}\op{a}_{B}+\op{S}^+\op{S}^-}\, \nonu \\
=& \frac{1}{3}\average{\alpha_{A}\beta_{A}+\alpha_{B}\beta_{B}+x}\, ,
}
\eqa{LiebNumfl}{
K =& \dfrac{\expect{(\dagop{a}_{A}\op{a}_{A}+\dagop{a}_{B}\op{a}_{B}+\op{S}^+\op{S}^-)^2}}{\expect{\dagop{a}_{A}\op{a}_{A}+\dagop{a}_{B}\op{a}_{B}+\op{S}^+\op{S}^-}} -3\nu \, \nonu \\
=& \dfrac{\average{\tilde{n}^2+2x\tilde{n}+x}}{\average{\tilde{n}+x}} -3\nu\, 
}
where we have introduced the shorthand $\tilde{n}=\alpha_A\beta_A+\alpha_B\beta_B$.  In obtaining the second expression for $K$, we have also used $(\op{S}^+\op{S}^-)^2=\op{S}^+\op{S}^-$, which both simplifies the estimator and avoids introducing unnecessary additional sampling noise associated with higher powers of $x$.

We also investigate correlations between the spins and bosons in each unit cell:
\eqy{SBcorr}{
C_{LX}=& \dfrac{\expect{\op{n}_L\op{S}^X}}{\expect{\op{n}_L}\expect{\op{S}^X}} =\dfrac{\average{\alpha_L\beta_L(w+z)}}{\average{\alpha_L\beta_L}\average{w+z}}\, , \nonu\\
C_{LY}=& \dfrac{\expect{\op{n}_L\op{S}^Y}}{\expect{\op{n}_L}\expect{\op{S}^Y}} =\dfrac{\average{\alpha_L\beta_L(w-z)}}{\average{\alpha_L\beta_L}\average{w-z}}\, , \nonu\\
C_{LZ}=& \dfrac{\expect{\op{n}_L\op{S}^Z}}{\expect{\op{n}_L}\expect{\op{S}^Z}} =\dfrac{\average{\alpha_L\beta_L(x-\frac{1}{2})}}{\average{\alpha_L\beta_L}\average{x-\frac{1}{2}}}\, ,
}
where $L\in\{A,B\}$ labels the two bosonic sites, and in all eqs. \eqn{LiebNume}-\eqn{SBcorr} we have suppressed the index $j$ that labels the unit cells, which is also averaged over wherever a stochastic average $\average{...}$ is calculated.  

\fig{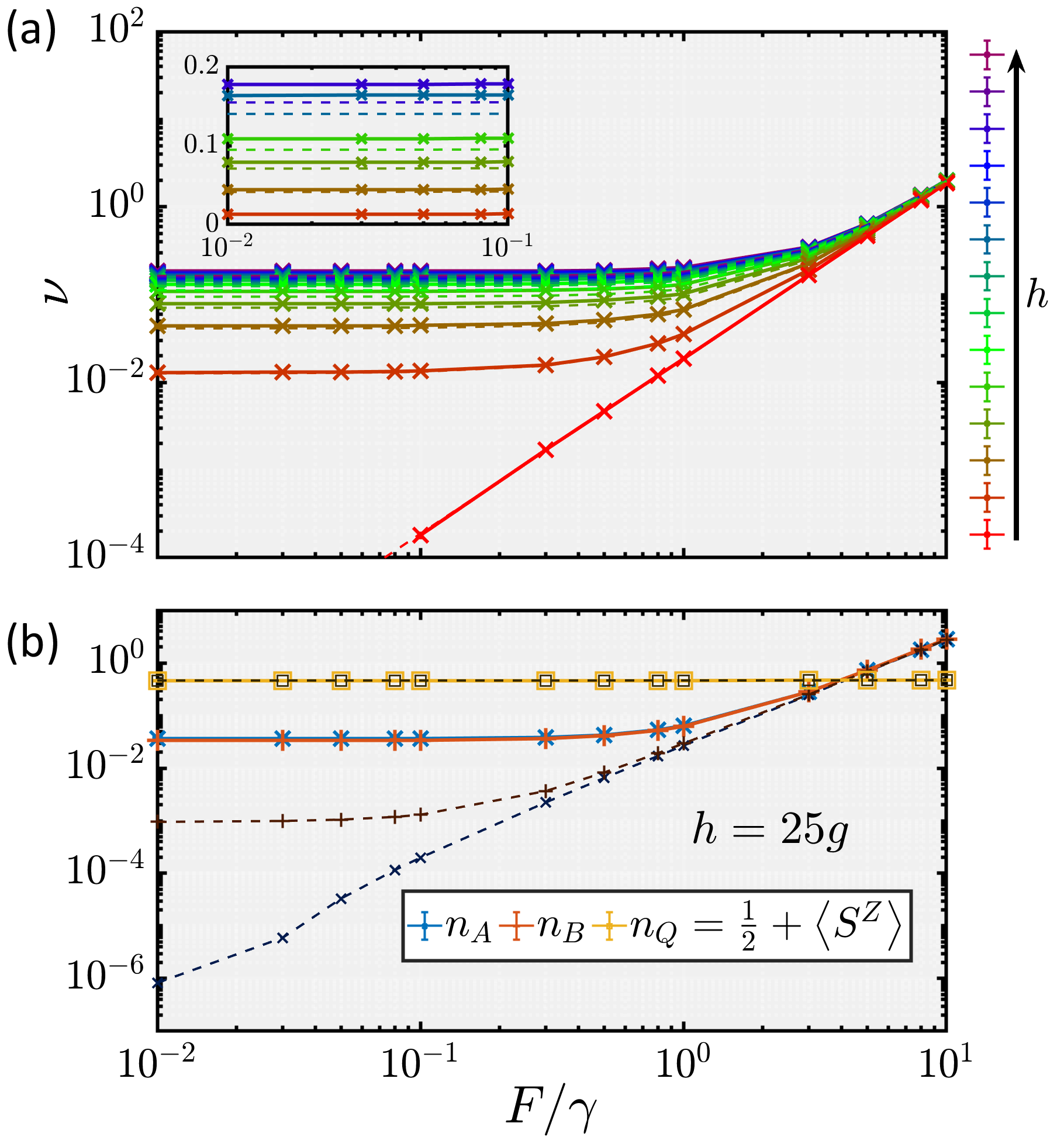}{{\bf 
Scaling of total excitations} with spin and boson drive strengths in a 20 unit cell {DDJCH} quasi-Lieb chain (40 cavity modes, 20 spins). a) \mbox{Average} \mbox{excitation} number $\nu$ versus coherent drive $F$ ({log-log} scale) for \mbox{$h={0,2,4,6,8,10,12,14,16,18,20,25,30,50}g$} (bottom to top). Dashed lines show mean-field solutions. \mbox{Inset:} linear-scale $\nu$ at small $F$ for selected $h={2,4,6,8,16,25}g$. \mbox{b) Spatially} averaged occupations $n_A$ (blue $\times$), $n_B$ (red +), and $n_Q$ (yellow squares) versus $F$ for $h=25g$. Darker dashed lines show the corresponding mean-field solutions. \label{JCHLiebExc}}

Equivalently to Biondi {\it et al.}, we study the behaviour over orders of magnitude in the bosonic coherent drive $F_A=F_B=F$.  To explore the influence of drive and dissipation on the spins in this model, we also vary the spin drive $h$.  Other parameters are $g=\gamma$, $J=3\gamma$, $\Gamma_\downarrow=10\gamma$, and detunings $\omega_C=\omega_0=0$.  We compare the results of our positive-P method to approximate mean field solutions, which in practice we find by solving the equations for $\expect{\op{a}}=\alpha_{MF}=\beta_{MF}^*$, $\expect{\op{S}^Z}=x_{MF}-\frac{1}{2}$, $\expect{\op{S}^-}=z_{MF}=w_{MF}^*$, produced by neglecting the noise terms in the Ito stochastic equations \eqn{Ito_full}, so as to see which observed behaviours can not be captured at the mean field level.  Notably, all normalised correlation functions, such as $g^{(2)}$ and the $C$ defined in equations \eqn{SBcorr} above, are 1 by definition in the mean field \mbox{approximation.}  

We begin with the average number of excitations $\nu$, shown in Fig.~\ref{JCHLiebExc}(a).  In the case without dissipation on the spins \cite{Biondi15}, $\nu$ saturates for intermediate values of the drive strength $F \sim \gamma$.  We can see that in our simulations with significant spin dissipation, this saturation no longer \mbox{occurs} with $F$, leading to a smooth power law dependence of $\nu\propto (F/\gamma)^2$ at $h=0$ and in the $F\gg\gamma$ regime.  For non-zero $h$, $\nu$ plateaus as $F\lesssim\gamma$; the $\nu$ value in this plateau seems to saturate for large values of $h$.  This \mbox{behaviour} is also captured qualitatively at the mean field level but deviations from the exact result provided by our \mbox{positive-P} method occur particularly at large $h$.  This suggests the accuracy of the mean field approach may begin to break down when the behaviour of the spins dominates the \mbox{dynamics.}  

To investigate this further, we look at the individual spatially averaged occupations of each mode, $n_A$, $n_B$, and $n_Q=\expect{\op{S}^+\op{S}^-} = \frac{1}{2}+\expect{\op{S}^Z}$, an example of which for $h = 25g$ is given in Fig.~\ref{JCHLiebExc}(b).  On the logarithmic scale, it can be seen that while the spin excitations $n_Q$ are \mbox{captured} well at the mean-field level, there are qualitative deviations in the behaviour of the bosons at low $F$: the mean-field solution tends towards a spatial pattern typical of the flat-band state in Lieb lattices, where the $A$ sites become ``dark'' with occupations tending to 0 compared to the $B$ sites.  However, when the quantum noise is included via the full positive-P solution, a greater incoherent population on both sites overwhelms this spatial pattern.  In this case this incoherent population is approximately equal on the $A$ and $B$ sites.  

Moving on to the correlations, we begin with the number fluctuations $K$, whose dependence on the two forms of driving are shown in Fig.~\ref{JCHLiebCorr}(a).  By definition, $K=0$ at the mean-field level, so any non-zero $K$ is a departure from that behaviour.  Let us observe the way the fluctuations vary upon changing each of the drives.  At the lower $F$, increasing $h$ results in a decrease of fluctuations from $K\sim1$ to a plateau at $K\approx0.5$ for large $h$; this \mbox{mirrors} the behaviour previously seen in the saturated state without spin dissipation \cite{Biondi15}.  Upon increasing $F$ in our case, there is a crossover to coherent behaviour signalled by a strong suppression of number fluctuations $K$, that occurs above $F\sim\gamma$.  The decrease in $K$ with $F$ appears to begin slightly earlier for smaller values of $h$.  

An example of the second-order correlations $g^{(2)}(0)$ of the bosonic modes is given in Fig.~\ref{JCHLiebCorr}(b) for $h=25g$.  We see that a crossover from incoherent, thermal-like \mbox{behaviour} $g^{(2)}(0)\gtrsim2$ to coherent behaviour \mbox{$g^{(2)}(0)\approx1$} occurs around $F\sim\gamma$.  The onset of this crossover in \mbox{increasing} $F$ occurs slightly earlier on the $A$ sites, which are directly coupled to the spins, compared to the $B$ sites.  An interesting behaviour, also shown in Fig.~\ref{JCHLiebCorr}(b), is seen in the corresponding spin-boson correlations $C$ \mbox{(eqn. \eqn{SBcorr}).}  Most notably, the crossover region is accompanied by a distinctive increase in $C_{AZ}$, which suggests that greater \mbox{correlations} between the directly coupled spin and \mbox{boson} modes may signal the intermediate state \mbox{between} coherent and incoherent behaviour.  The other $C$ for non-zero projections of the spin are much closer to 1 (uncorrelated), but can still show some interesting \mbox{qualitative} \mbox{dependence} on $F$, such as a shift from \mbox{correlated} $C>1$ to anti-correlated $C<1$ in $C_{AY}$. 

Overall, while the ``incompressibility'' behaviour with increasing $F$ observed in ref.~\cite{Biondi15} is destroyed when excitations are allowed to escape through the spin modes due to the presence of strong spin decay, similar behaviour \mbox{returns} at small $F$ when the spin decay is compensated by a coherent driving $h$ of the spins directly, leading to non-zero excitations of the spin modes in the steady state.  Crucially, this example demonstrates our method is able to simulate large DDJCH lattices in regimes where the behaviour differs qualitatively from purely mean-field predictions.  

\fig{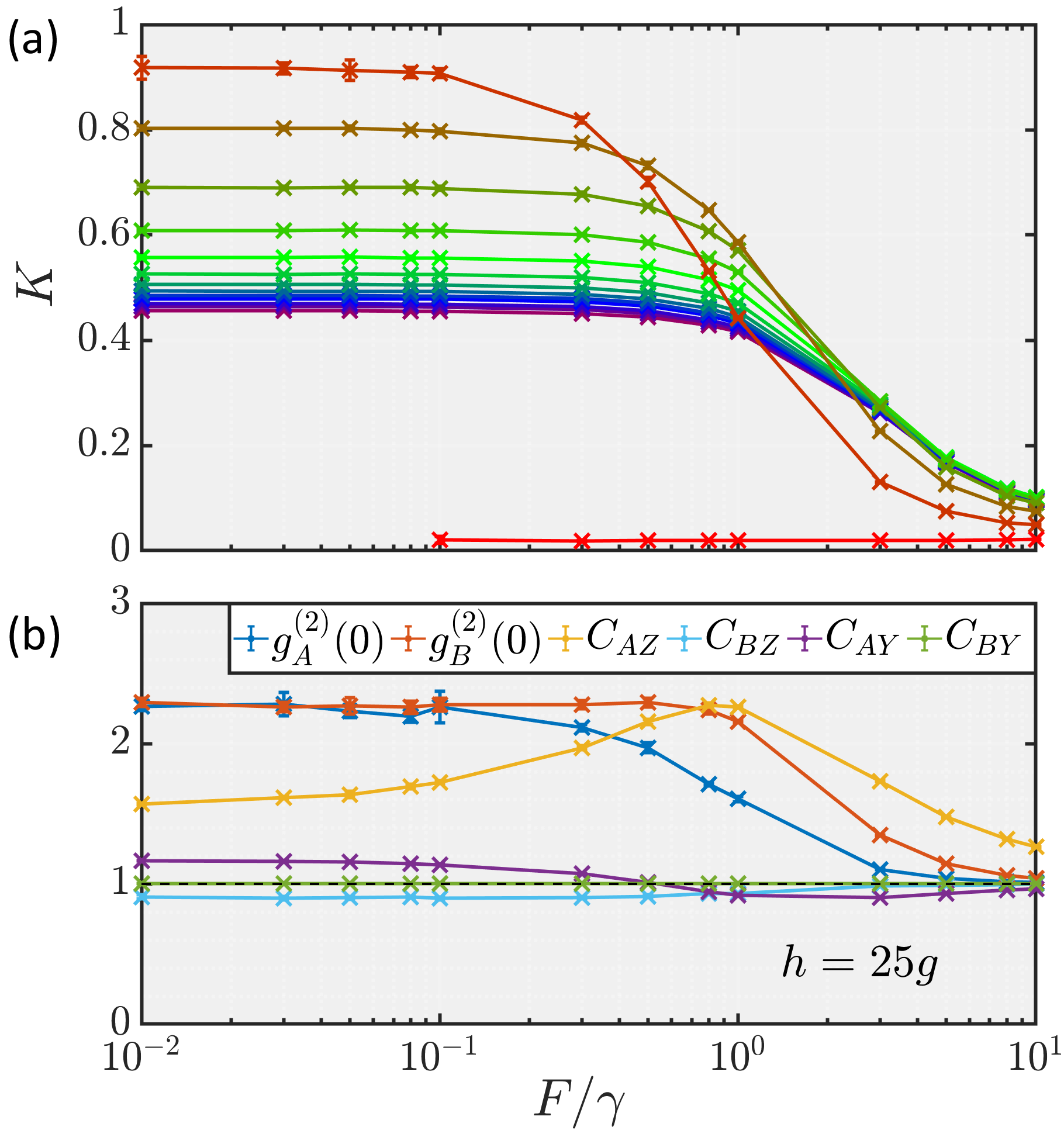}{{\bf 
Variation of correlations with spin and \mbox{boson} drive strengths.} a) Number fluctuations $K$ versus \mbox{coherent} drive $F$ for spin drives $h$, coloured as in Fig.~\ref{JCHLiebExc}(a); \mbox{$h=0$,} \mbox{$F<0.1\gamma$} is omitted due to poor signal-to-noise from \mbox{vanishing} occupations. b) $g^{(2)}_A(0)$ (blue), $g^{(2)}_B(0)$ (red), and spin-boson correlations $C_{AY}$ (purple), $C_{AZ}$ (yellow), $C_{BY}$ (green), and $C_{BZ}$ (cyan) versus $F$ for $h=25g$, averaged over trajectories and unit cells. The dashed black line marks the mean-field/coherent-state value $C_{\rm MF}=g^{(2)}_{\rm MF}(0)\equiv1$. $C_{AX}$ and $C_{BX}$ are omitted because $\expect{\op{S}^X}\approx0$ gives vanishing normalisation, resulting in divergences and poor signal-to-noise.  \label{JCHLiebCorr}}

\section*{Discussion}

We have introduced a new GDM kernel approach that enables scalable and exact stochastic simulations of \mbox{spin-$\frac{1}{2}$} degrees of freedom in regimes where dissipation is sufficiently strong to counteract noise self-amplification, extending to spins the advantages that the established positive-P method offers for purely bosonic systems \cite{PRXQuantum.2.010319}. This stands in contrast to previous attempts to construct positive-P representations for spins, which can develop systematic simulation errors without an accompanying noise instability to provide a clear warning of failure \cite{Mandt15,Chuchurka24b,Leonau25arxiv}. Here, we have focused on coupled spin-boson systems, including the single-site DDJC and lattice DDJCH models, motivated in particular by the seamless integration of the GDM representation with the conventional positive-P treatment of bosonic modes. For such hybrid systems, the method offers an additional advantage: diffusion gauges allow noise to be redistributed between the spin and bosonic sectors, providing some control over the stability boundary and thereby extending the \mbox{accessible} parameter regime. Recent developments \mbox{using} machine learning to optimise stochastic gauges and \mbox{enlarge} the stable regimes of bosonic positive-P simulations \cite{cao2026neuralgaugeprepresentationopen} could therefore offer a promising route to \mbox{further} improve the GDM positive-P approach.

More broadly, moving beyond the conventional restriction to local kernels constructed solely from bra and ket states opens a substantially wider space of possible phase-space representations. Such constructions have rarely been explored, with a notable exception being the density-matrix-element phase-space methods developed by the Hamburg group \cite{Chuchurka23arxiv,Chuchurka24,Chuchurka24b,Leonau25arxiv}. The framework developed here therefore suggests several immediate \mbox{directions} for extension. It can be applied naturally to a broad class of dissipative spin-boson models, including Rabi, Dicke, and Tavis-Cummings systems, which have modified couplings compared to the JC, or multiple spins interacting with a common bosonic mode.

An especially promising direction is the application of the method to dissipative spin lattices without bosonic degrees of freedom. In such systems, we expect \mbox{spin-spin} interactions between different sites to generate the noise terms, similarly to in SU(2) state-based \mbox{approaches} \cite{Ng13}. A key attraction of stochastic phase-space \mbox{methods} is that their computational structure is largely insensitive to system geometry and dimensionality. This could make the present framework particularly valuable for systems with long-range interactions, which remain challenging for \mbox{numerical} methods otherwise well suited to spin lattices, including tensor-network approaches \cite{dunham2025,hryniuk2026}. \mbox{Understanding} how dissipation competes with the noise generated by short- and long-range spin-spin interactions will therefore be central to determining the \mbox{ultimate} regime of \mbox{validity} of the method.
Taken \mbox{together,} these results establish the GDM \mbox{positive-P} framework as a flexible route toward stochastic simulation of large \mbox{interacting} spin and spin-boson systems \mbox{beyond} the reach of direct density-matrix methods. The combination of exact phase-space dynamics, compatibility with bosonic positive-P techniques, and the freedom to engineer both kernels and stochastic gauges opens a broad range of opportunities for treating driven dissipative quantum many-body systems that remain \mbox{difficult} to \mbox{access} with existing numerical approaches.

\section*{Methods}
\subsection{Stochastic equations for the GDM kernel}
\label{sec:GDMSDE}

Here we outline how stochastic differential equations for solving the DDJCH model are obtained when using the GDM kernel operator for the spins.  

\

\subsubsection{Obtaining the Fokker-Planck equation}
\label{sec:FPE}

Following the usual recipe for the derivation of the positive-P method, with the kernel operator $\op{\Lambda}_S$ defined by \eqn{xkern}, the next step is to find correspondences between spin operators and differential operators acting on $\op{\Lambda}_S$.  While normally for the bosonic or SU(2) coherent state kernels this is done by looking at the action of operators on the coherent states, here we directly compare $2\times2$ matrix multiplication of \eqn{xkern} with various spin operators to the elements formed from its first derivatives.  This results in the following relations:
\begin{widetext}
\eqy{xopcon}{
&\op{S}^Z\op{\Lambda}_S = \left[x-\frac{1}{2}+x(1-x)\dada{}{x} +z(1-x)\dada{}{z} -wx\dada{}{w}\right]\op{\Lambda}_S\, , \nonu\\
&\op{S}^+\op{\Lambda}_S = \left[w+w(1-x)\dada{}{x} +(1-x-zw)\dada{}{z} -w^2\dada{}{w}\right]\op{\Lambda}_S\, , \nonu\\
&\op{S}^-\op{\Lambda}_S = \left[z-zx\dada{}{x} -z^2\dada{}{z} +(x-zw)\dada{}{w}\right]\op{\Lambda}_S\, , \nonu\\
&\op{\Lambda}_S\op{S}^Z = \left[x-\frac{1}{2}+x(1-x)\dada{}{x} -zx\dada{}{z} +w(1-x)\dada{}{w}\right]\op{\Lambda}_S\, , \nonu\\
&\op{\Lambda}_S\op{S}^+ = \left[w-wx\dada{}{x} +(x-zw)\dada{}{z} -w^2\dada{}{w}\right]\op{\Lambda}_S\, , \nonu\\
&\op{\Lambda}_S\op{S}^- = \left[z+z(1-x)\dada{}{x} -z^2\dada{}{z} +(1-x-zw)\dada{}{w}\right]\op{\Lambda}_S\, , \nonu\\
&\op{S}^Z\op{\Lambda}_S\op{S}^Z = \left[\frac{1}{4} - \frac{z}{2}\dada{}{z} - \frac{w}{2}\dada{}{w}\right]\op{\Lambda}_S\, . 
}
\end{widetext}
The final of these relations shows that the result of action by multiple spin operators can also often be simplified beyond just applying the correspondences for each one sequentially.  

Applying these relations, along with those for the bosonic modes, to the master equation \eqn{ME}, and \mbox{integrating} by parts converts it to a FPE for the \mbox{distribution $P(\vec{v})$} 
\eq{FPE}{
\dada{P}{t} = -\dada{}{v_m}\left[A_m P\right] +\frac{1}{2}\frac{\partial^2}{\partial v_m \partial v_{m'}}\left[ D_{m,m'}P\right]\, , 
}
with summation over repeated indices.  In the basis $(z_j, w_j, x_j, \alpha_j, \beta_j)$, the drift has components for phase space variables on site $j$ 
\eq{drift}{
A_j = \begin{pmatrix} 
(-i\omega_0-\frac{\Gamma_\phi}{2}-\frac{\Gamma_\downarrow}{2})z_j -i(\frac{h}{2}+g\alpha_j)(1-2x_j) \\
(i\omega_0-\frac{\Gamma_\phi}{2}-\frac{\Gamma_\downarrow}{2})w_j +i(\frac{h}{2}+g\beta_j)(1-2x_j) \\	 
\frac{ih}{2}(z_j-w_j)+ig(z_j\beta_j-w_j\alpha_j) -\Gamma_\downarrow x_j \\
(-i\omega_C-\frac{\gamma}{2})\alpha_j -igz_j -iF +i\sum_{j'}J_{j,j'}\alpha_{j'}\\
(i\omega_C-\frac{\gamma}{2})\beta_j +igw_j +iF -i\sum_{j'}J_{j,j'}\beta_{j'}
\end{pmatrix}\, , 
}
while the diffusion matrix is block diagonal in local \mbox{submatrices} for each site $j$, which have components 
\eq{diffusion}{
D_j = ig\begin{pmatrix} 
0 & 0 & 0 & z_j^2 & \epsilon_j \\
0 & 0 & 0 & -\epsilon_j & -w_j^2 \\
0 & 0 & 0 & z_jx_j & -w_jx_j \\
z_j^2 & -\epsilon_j & z_jx_j & 0 & 0 \\
\epsilon_j & -w_j^2 & -w_jx_j & 0 & 0 
\end{pmatrix}\, ,
}
where we have used a shorthand notation $\epsilon_j \equiv x_j-z_jw_j$.  
The Ito stochastic differential equations for trajectories of the phase space variables $\vec{v}$ that result from this are then 
\eq{SDE}{
\dada{\vec{v}}{t} = \vec{A} + B\vec{\xi}\, ,
}
where $\vec{\xi}$ are a set of uncorrelated real Gaussian noises $\expect{\xi_m(t)\xi_{m'}(t')}=\delta_{m,m'}\delta(t-t')$ and the noise matrix $B$ is chosen such that $D = BB^T$.

\subsubsection{Diffusion Gauges}
\label{sec:DiffuG}

The decomposition of the diffusion matrix $D$ into noise matrices $B$ is, in general, not unique.  It can be seen from the definition that any valid solution for $B$ can be \mbox{converted} to a new solution $B' = BO$ when right \mbox{multiplying} by an orthogonal matrix $OO^T = \mathbb{I}$.  It has been noted in previous work that particular choices made with this freedom, known as diffusion gauges, can \mbox{sometimes} \mbox{improve} the stability of the positive-P method and in a manner that is scalable \cite{Deuar06b}.  In the case of the model being \mbox{studied} here, there is a clear physical \mbox{motivation} to guide our choice of diffusion gauge: since the diffusion terms \mbox{couple} the spin and boson modes, the diffusion gauge can be used to shift the balance of noise between the spins and bosons.  A broad class of forms for the noise matrix (\mbox{elements} for site $j$) that demonstrate this is given by 
\begin{widetext}
\eq{diffG}{
B_j =  \sqrt{\frac{ig}{2}}\begin{pmatrix}
z_j/\sqrt{G} & iz_j/\sqrt{G} & z_j/\sqrt{G} & iz_j/\sqrt{G} & 0 & 0 & 1/\sqrt{G} & i/\sqrt{G} \\
w_j/\sqrt{G} & iw_j/\sqrt{G} & w_j/\sqrt{G} & iw_j/\sqrt{G} & 1/\sqrt{G} & i/\sqrt{G} & 0 & 0 \\
x_j/\sqrt{G} & ix_j/\sqrt{G} &  x_j/\sqrt{G} & ix_j/\sqrt{G} & 0 & 0 & 0 & 0 \\
z_j\sqrt{G} & -iz_j\sqrt{G} & 0 & 0 & -x_j\sqrt{G} & ix_j\sqrt{G} & 0 & 0 \\
0 & 0 & -w_j\sqrt{G} & iw_j\sqrt{G} & 0 & 0 & x_j\sqrt{G} & -ix_j\sqrt{G} 
\end{pmatrix}\, ,
}
\end{widetext}
where the variable $G$ sets the diffusion gauge.  Its \mbox{effect} is to make the standard deviation of spin noise $G$ times smaller relative to boson noise, in comparison to the \mbox{balance} in the un-gauged ($G = 1$) equations.  While for the results presented here we consider only \mbox{constant} values of $G$, it can in principle be any arbitrary \mbox{function} of the phase space variables, time, $j$, and other \mbox{parameters.}  Substituting equations \eqn{drift} and \eqn{diffG} into the \mbox{corresponding} terms in equation \eqn{SDE}, we get the full set of Ito stochastic differential equations: 
\eqa{Ito_full}{
\nonu \\
\dada{z_j}{t} =&\, (-i\omega_0-\frac{\Gamma_\phi}{2}-\frac{\Gamma_\downarrow}{2})z_j -i(\frac{h}{2}+g\alpha_j)(1-2x_j) \nonu 
\\& +\sqrt{\frac{ig}{G}}(z_j\eta_{Z,j} +z_j\eta_{W,j} +\eta_{B,j})\, , \nonu\\ 
\nonu \\
\dada{w_j}{t} =&\, (i\omega_0-\frac{\Gamma_\phi}{2}-\frac{\Gamma_\downarrow}{2})w_j +i(\frac{h}{2}+g\beta_j)(1-2x_j) \nonu 
\\& +\sqrt{\frac{ig}{G}}(w_j\eta_{Z,j} +w_j\eta_{W,j} +\eta_{A,j})\, , \nonu\\ 
\nonu \\
\dada{x_j}{t} =&\, \frac{ih}{2}(z_j-w_j)+ig(z_j\beta_j-w_j\alpha_j) -\Gamma_\downarrow x_j \nonu 
\\& +\sqrt{\frac{ig}{G}}x_j(\eta_{Z,j} +\eta_{W,j})\, , \nonu\\
\nonu \\
\dada{\alpha_j}{t} =&\, (-i\omega_C-\frac{\gamma}{2})\alpha_j -igz_j -iF +i\sum_{j'}J_{j,j'}\alpha_{j'} \nonu 
\\& +\sqrt{igG}(z_j\eta^*_{Z,j} -x_j\eta^*_{A,j})\, , \nonu\\
\nonu \\
\dada{\beta_j}{t} =&\, (i\omega_C-\frac{\gamma}{2})\beta_j +igw_j +iF -i\sum_{j'}J_{j,j'}\beta_{j'} \nonu 
\\& -\sqrt{igG}(w_j\eta^*_{W,j} -x_j\eta^*_{B,j})\, ,
}
where $\eta_{X\in\{Z,W,A,B\},j}$ are uncorrelated complex Gaussian noises with $\expect{\eta_{X,j}(t)\eta^*_{X',j'}(t')} = \delta_{X,X'}\delta_{j,j'}\delta(t-t')$ that result from the combinations of real noises implied by the form of $B_j$, e.g.~$\eta_{Z,j}=(\xi_{1,j}+i\xi_{2,j})/\sqrt{2}$.  The Stratonovich form for these equations, needed for some numerical \mbox{integration} algorithms, is given in section~\ref{StrC}.  For the simulation results presented in this work, we \mbox{numerically} solve these stochastic differential equations using the xmds2 software \cite{XMDS_CPhysComm.184.201}.  

\subsection{Investigations into the Schwinger Boson Approach}
\label{Schwing}

This section will briefly explain our investigations on using a Schwinger boson mapping of the spins to treat all degrees of freedom using the standard bosonic \mbox{positive-P} kernel operator (\ref{bosonLambda}), and the limitations this approach encounters that ultimately discount it from being useful for modelling the sorts of physical systems we consider in this work.  This approach for treating spins within a \mbox{positive-P} framework had been initially implemented for closed spin chains \cite{Ng11}, where as usual for \mbox{positive-P} methods in closed systems, it allows for simulations up to a finite time until the noise instability occurs.  

The Schwinger boson mapping involves converting each spin to a pair of new bosonic modes, with annihilation operators we label as $\op{b}_{\uparrow,j},\op{b}_{\downarrow,j}$, and which we will ultimately assign the corresponding phase space variables $\alpha_{\uparrow,j},\beta_{\uparrow,j}$ and $\alpha_{\downarrow,j},\beta_{\downarrow,j}$, respectively.  The spin operators are transformed as: 
\eqy{SchwingMap}{
&\op{S}^Z_j = \frac{1}{2}(\dagop{b}_{\uparrow,j}\op{b}_{\uparrow,j} - \dagop{b}_{\downarrow,j}\op{b}_{\downarrow,j})\, , \nonu\\ 
&\op{S}^+_j = \dagop{b}_{\uparrow,j}\op{b}_{\downarrow,j}\, , \nonu\\
&\op{S}^-_j = \dagop{b}_{\downarrow,j}\op{b}_{\uparrow,j}\, ,
}
with an additional constraint on allowed states
\eq{SchwingConst}{\dagop{b}_{\uparrow,j}\op{b}_{\uparrow,j} + \dagop{b}_{\downarrow,j}\op{b}_{\downarrow,j} = \mathbb{I} \, ,}
which preserves the total spin length.  This transforms the local hamiltonian \eqn{H_JC}, for example, as 
\eqa{H_Schwing}{
\op{H}^{JC}_j =&\, \omega_C\dagop{a}_j\op{a}_j +\frac{\omega_0}{2}\left(\dagop{b}_{\uparrow,j}\op{b}_{\uparrow,j} - \dagop{b}_{\downarrow,j}\op{b}_{\downarrow,j}\right) +F\left(\op{a}_j+\dagop{a}_j\right) \nonu \\ 
&+g\left(\dagop{b}_{\uparrow,j}\op{b}_{\downarrow,j}\op{a}_j +\dagop{a}_j\dagop{b}_{\downarrow,j}\op{b}_{\uparrow,j}\right) \nonu \\ &+\frac{h}{2}\left(\dagop{b}_{\uparrow,j}\op{b}_{\downarrow,j}+\dagop{b}_{\downarrow,j}\op{b}_{\uparrow,j}\right)\, .   
}
Applying the process of converting the master equation to a FPE, as outlined in section~\ref{sec:GDMSDE}, finally leads in this framework to a set of $6j$ coupled stochastic differential equations for the phase space trajectories of both actual and Schwinger bosons, which we write in the Stratonovich form (see section~\ref{StrC}) for our numerical simulations as: \pagebreak
\eqa{SchwingStrat}{
\dada{\alpha_{\uparrow,j}}{t} =&\, \left[\frac{-i(\omega_0-\sqrt{\Gamma_\phi}\xi_{\phi,j})}{2} -\frac{\Gamma_\downarrow}{2}(1+\alpha_{\downarrow,j}\beta_{\downarrow,j})\right]\alpha_{\uparrow,j} \nonu 
\\& -i\left(g\alpha_j+\frac{h}{2}\right)\alpha_{\downarrow,j} -\sqrt{\frac{\Gamma_\downarrow}{2}}\alpha_{\downarrow,j}\zeta_{A,j}\, , \nonu\\
\dada{\alpha_{\downarrow,j}}{t} =&\, \left[\frac{i(\omega_0-\sqrt{\Gamma_\phi}\xi_{\phi,j})}{2} -\frac{\Gamma_\downarrow}{2}\alpha_{\uparrow,j}\beta_{\uparrow,j}\right]\alpha_{\downarrow,j}  \nonu 
\\& -i\left(g\beta_j+\frac{h}{2}\right)\alpha_{\uparrow,j} -\sqrt{ig}\alpha_{\uparrow,j}\eta_{A,j} \nonu 
\\& +\sqrt{\frac{\Gamma_\downarrow}{2}}\alpha_{\uparrow,j}(\zeta_{A,j}^* +\zeta_{B,j})\, , \nonu\\
\dada{\alpha_j}{t} =&\, (-i\omega_C-\frac{\gamma}{2})\alpha_j -ig\alpha_{\uparrow,j}\beta_{\downarrow,j} +\sqrt{ig}\eta^*_{A,j} \nonu 
\\& -iF +i\sum_{j'}J_{j,j'}\alpha_{j'}\, , \nonu\\
\dada{\beta_{\uparrow,j}}{t} =&\, \left[\frac{i(\omega_0-\sqrt{\Gamma_\phi}\xi_{\phi,j})}{2} -\frac{\Gamma_\downarrow}{2}(1+\alpha_{\downarrow,j}\beta_{\downarrow,j})\right]\beta_{\uparrow,j} \nonu 
\\& +i\left(g\beta_j+\frac{h}{2}\right)\beta_{\downarrow,j} -\sqrt{\frac{\Gamma_\downarrow}{2}}\beta_{\downarrow,j}\zeta_{B,j}\, , \nonu\\
\dada{\beta_{\downarrow,j}}{t} =&\, \left[\frac{-i(\omega_0-\sqrt{\Gamma_\phi}\xi_{\phi,j})}{2} -\frac{\Gamma_\downarrow}{2}\alpha_{\uparrow,j}\beta_{\uparrow,j}\right]\beta_{\downarrow,j}  \nonu 
\\& +i\left(g\alpha_j+\frac{h}{2}\right)\beta_{\uparrow,j} +\sqrt{ig}\beta_{\uparrow,j}\eta_{B,j} \nonu 
\\& +\sqrt{\frac{\Gamma_\downarrow}{2}}\beta_{\uparrow,j}(\zeta_{A,j} +\zeta_{B,j}^*)\, , \nonu\\
\dada{\beta_j}{t} =&\, (i\omega_C-\frac{\gamma}{2})\beta_j +ig\beta_{\uparrow,j}\alpha_{\downarrow,j} +\sqrt{ig}\eta^*_{B,j} \nonu 
\\& +iF -i\sum_{j'}J_{j,j'}\beta_{j'}\, ,
}
where $\xi_{\phi,j}$ are uncorrelated real gaussian noises with $\expect{\xi_{\phi,j}(t)\xi_{\phi,j'}(t')} = \delta_{j,j'}\delta(t-t')$, and both $\eta_{X\in\{A,B\},j}$ and $\zeta_{X\in\{A,B\},j}$ represent uncorrelated complex gaussian noises with \mbox{$\expect{\eta_{X,j}(t)\eta^*_{X',j'}(t')} = \delta_{X,X'}\delta_{j,j'}\delta(t-t')$} and equivalently for $\zeta_{X\in\{A,B\},j}$.  If we consider the observable corresponding to the constraint \eqn{SchwingConst}, \mbox{$\expect{\tilde{\mathbb{I}}} = \expect{\dagop{b}_{\uparrow,j}\op{b}_{\uparrow,j} + \dagop{b}_{\downarrow,j}\op{b}_{\downarrow,j}} = \average{\alpha_{\uparrow,j}\beta_{\uparrow,j}+\alpha_{\downarrow,j}\beta_{\downarrow,j}}$,} the drift of these equations conserves the spin length exactly, by construction, but the noise terms ultimately may only conserve it on-average over many trajectories, meaning individual trajectories may cross into parts of the phase space for the Schwinger bosons that no longer represent spins of fixed length.  

Moreover, a second problem is that coherent state \mbox{initial} conditions, convenient for the stochastic simulation, involve a spread of boson number around the mean. This makes them not correspond exactly to a desired spin length, especially when the spin length is small.  \mbox{Initial} conditions that \emph{do} correspond exactly to Fock states are known \cite{Olsen09} but involve a rather broad distribution that pumps a great amount of additional noise into the \mbox{system,} which also grows with system size.  

\fig{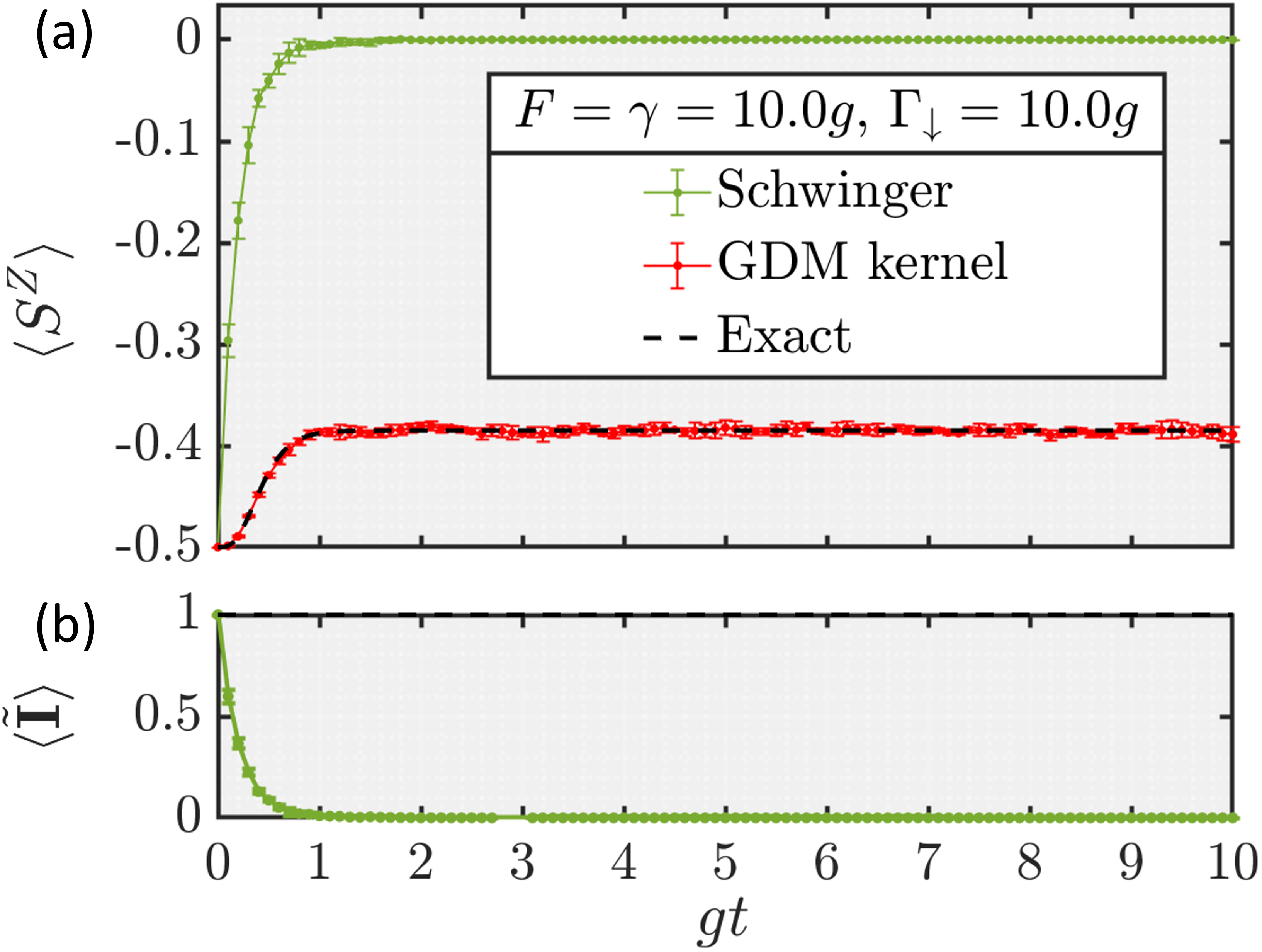}{{\bf Errors in Schwinger boson \mbox{positive-P} method with spin decay.}  a) Comparison of $\expect{\op{S}^Z}$ for the GDM spin kernel and Schwinger boson based \mbox{positive-P} \mbox{methods} to the exact solution for a single site DDJC model with $F = \gamma = \Gamma_\downarrow = 10g$, all other parameters $=0$; unlike the GDM kernel, the Schwinger boson method drastically fails to match exact behaviour.  b) Average of the Schwinger boson constraint $\expect{\tilde{\mathbb{I}}}$ from the Schwinger boson positive-P simulation.  This quantity should remain unity if the spin length is properly \mbox{preserved,} but instead \mbox{decays to 0.}  \label{SchwingError}}

We show in Fig.~\ref{SchwingError} that the Schwinger boson based method fails to reproduce the correct behaviour when spin decay is included.  It appears that in these cases the measure of the constraint $\expect{\tilde{\mathbb{I}}}$ fails to be preserved, and instead rapidly decays to zero.  Since without spin decay, the simulations eventually suffer the usual noise \mbox{instability,} limiting the useful simulations time, but with decay they fail to give accurate results at all, we \mbox{ultimately} concluded that a positive-P method for spins based on the Schwinger boson mapping was not a useful way to proceed for this work.

\subsection{Stochastic equations for the SU(2) kernel}
\label{SESU2}

Here, we will elaborate on our investigations using the SU(2) coherent state kernel.  The following relations \mbox{result} from applying spin operators to the SU(2) kernel operator \eqn{SU2k}:
\eqy{SU2opcon}{
&\op{S}^Z\op{\Lambda}_S = [z\dada{}{z} -\frac{1}{2}]\op{\Lambda}_S\, , \nonu\\
&\op{S}^+\op{\Lambda}_S = \dada{}{z} \op{\Lambda}_S\, , \nonu\\
&\op{S}^-\op{\Lambda}_S = z[1 -z\dada{}{z}]\op{\Lambda}_S\, , \nonu\\
&\op{\Lambda}_S\op{S}^Z = [w\dada{}{w} -\frac{1}{2}]\op{\Lambda}_S\, , \nonu\\
&\op{\Lambda}_S\op{S}^+ = w[1 -w\dada{}{w}]\op{\Lambda}_S\, , \nonu\\
&\op{\Lambda}_S\op{S}^- = \dada{}{w}\op{\Lambda}_S\, .
}
Applying these to the master equation gives a FPE with drift components in the $(z_j, w_j, \alpha_j, \beta_j)$ basis of
\eqa{driftSU2}{
\lefteqn{A_j =}&& \\
\nonu&&\begin{pmatrix} 
\tilde{\Gamma}^-_jz_j -i\frac{h}{2}(1-z_j^2)-ig(\alpha_j-\beta_j z_j^2) \\
\tilde{\Gamma}^+_jw_j +i\frac{h}{2}(1-w_j^2)+ig(\beta_j-\alpha_j w_j^2) \\	 
(-i\omega_C-\frac{\gamma}{2})\alpha_j -ig\frac{z_j}{1+z_jw_j} -iF +i\sum_{j'}J_{j,j'}\alpha_{j'}\\
(i\omega_C-\frac{\gamma}{2})\beta_j +ig\frac{w_j}{1+z_jw_j} +iF -i\sum_{j'}J_{j,j'}\beta_{j'}
\end{pmatrix}\, ,
}
with shorthand $\tilde{\Gamma}^\pm_j=(\pm i\omega_0-\frac{\Gamma_\phi}{2}-\frac{\Gamma_\downarrow}{2}\frac{1+3z_j w_j}{1+z_j w_j})$.  The noise in this case comes from three different sources: the JC interaction $g$ generates noise like with the GDM kernel, but both forms of spin dissipation also contribute terms to the noise.  We can write the total diffusion matrix as contributions from each of these $D = D_g + D_\downarrow + D_\phi$ with: 
\eqy{diffSU2}{
&D_{g,j} = ig\begin{pmatrix} 
0 & 0 & z_j^2 & 0 \\
0 & 0 & 0 & -w_j^2 \\
z_j^2 & 0 & 0 & 0 \\
0 & -w_j^2 & 0 & 0
\end{pmatrix}\, , \nonu\\
&D_{\downarrow,j} = \Gamma_\downarrow\begin{pmatrix}
0 & z_j^2w_j^2 \\
z_j^2w_j^2 & 0
\end{pmatrix}\, , \nonu\\
&D_{\phi,j} = \Gamma_\phi\begin{pmatrix}
-z_j^2 & z_jw_j \\
z_jw_j & -w_j^2
\end{pmatrix}\, , 
} 
where $D_\downarrow$ and $D_\phi$ components are written in the $(z_j, w_j)$ basis, since they have no non-zero components for the bosonic parts.  We decompose these into independent noise matrices as 
\eqy{noiseSU2}{
&B_{g,j} =\sqrt{\frac{ig}{2}}\begin{pmatrix} 
z_j^2 & iz_j^2 & 0 & 0 \\
0 & 0 & iw_j^2 & -w_j^2 \\
1 & -i & 0 & 0 \\
0 & 0 & i & 1
\end{pmatrix}\, , \nonu\\
&B_{\downarrow,j} = \sqrt{\frac{\Gamma_\downarrow}{2}}\begin{pmatrix}
z_j^2 & iz_j^2 \\
w_j^2 & -iw_j^2
\end{pmatrix}\, , \nonu\\
&B_{\phi,j} = \sqrt{\Gamma_\phi}\begin{pmatrix}
iz_j \\
-iw_j
\end{pmatrix}\, , 
}
leading to the Ito stochastic differential equations: 
\eqy{ItoSU2}{
\dada{z_j}{t} =&\, \left[-i\omega_0-\frac{\Gamma_\phi}{2}+i\sqrt{\Gamma_\phi}\xi_{\phi,j}-\frac{\Gamma_\downarrow}{2}\frac{1+3z_j w_j}{1+z_j w_j}\right]z_j \nonu 
\\& +\sqrt{\Gamma_\downarrow}z_j^2\eta_{\downarrow,j} -i\frac{h}{2}(1-z_j^2) \nonu 
\\& -ig(\alpha_j-\beta_j z_j^2) +z_j^2\sqrt{ig}\eta_{Z,j} \, , \nonu\\
\nonu \\
\dada{w_j}{t} =&\, \left[i\omega_0-\frac{\Gamma_\phi}{2}-i\sqrt{\Gamma_\phi}\xi_{\phi,j}-\frac{\Gamma_\downarrow}{2}\frac{1+3z_j w_j}{1+z_j w_j}\right]w_j \nonu 
\\& +\sqrt{\Gamma_\downarrow}w_j^2\eta_{\downarrow,j}^* +i\frac{h}{2}(1-w_j^2) \nonu 
\\& +ig(\beta_j-\alpha_j w_j^2) -w_j^2\sqrt{ig}\eta_{W,j} \, , \nonu\\
\nonu \\
\dada{\alpha_j}{t} =&\, (-i\omega_C-\frac{\gamma}{2})\alpha_j -ig\frac{z_j}{1+z_jw_j} \nonu 
\\& -iF +i\sum_{j'}J_{j,j'}\alpha_{j'} +\sqrt{ig}\eta^*_{Z,j}\, , \nonu\\
\nonu \\
\dada{\beta_j}{t} =&\, (i\omega_C-\frac{\gamma}{2})\beta_j +ig\frac{w_j}{1+z_jw_j} \nonu 
\\& +iF -i\sum_{j'}J_{j,j'}\beta_{j'} +\sqrt{ig}\eta^*_{W,j}\, ,
}
where $\xi_{\phi,j}$ are uncorrelated real gaussian noises with $\expect{\xi_{\phi,j}(t)\xi_{\phi,j'}(t')} = \delta_{j,j'}\delta(t-t')$, and $\eta_{X\in\{Z,W,\downarrow\},j}$ are uncorrelated complex gaussian noises with $\expect{\eta_{X,j}(t)\eta^*_{X',j'}(t')} = \delta_{X,X'}\delta_{j,j'}\delta(t-t')$ as before.  Spin observables are calculated as
\eqy{SU2Obs}{
\expect{\op{S}^Z_j} = \frac{1}{2}\average{\frac{z_j w_j -1}{1+z_j w_j}}\, , \nonu\\
\expect{\op{S}^-_j} = \average{\frac{z_j}{1+z_j w_j}}\, , \nonu\\
\expect{\op{S}^+_j} = \average{\frac{w_j}{1+z_j w_j}}\, .
}

\figv{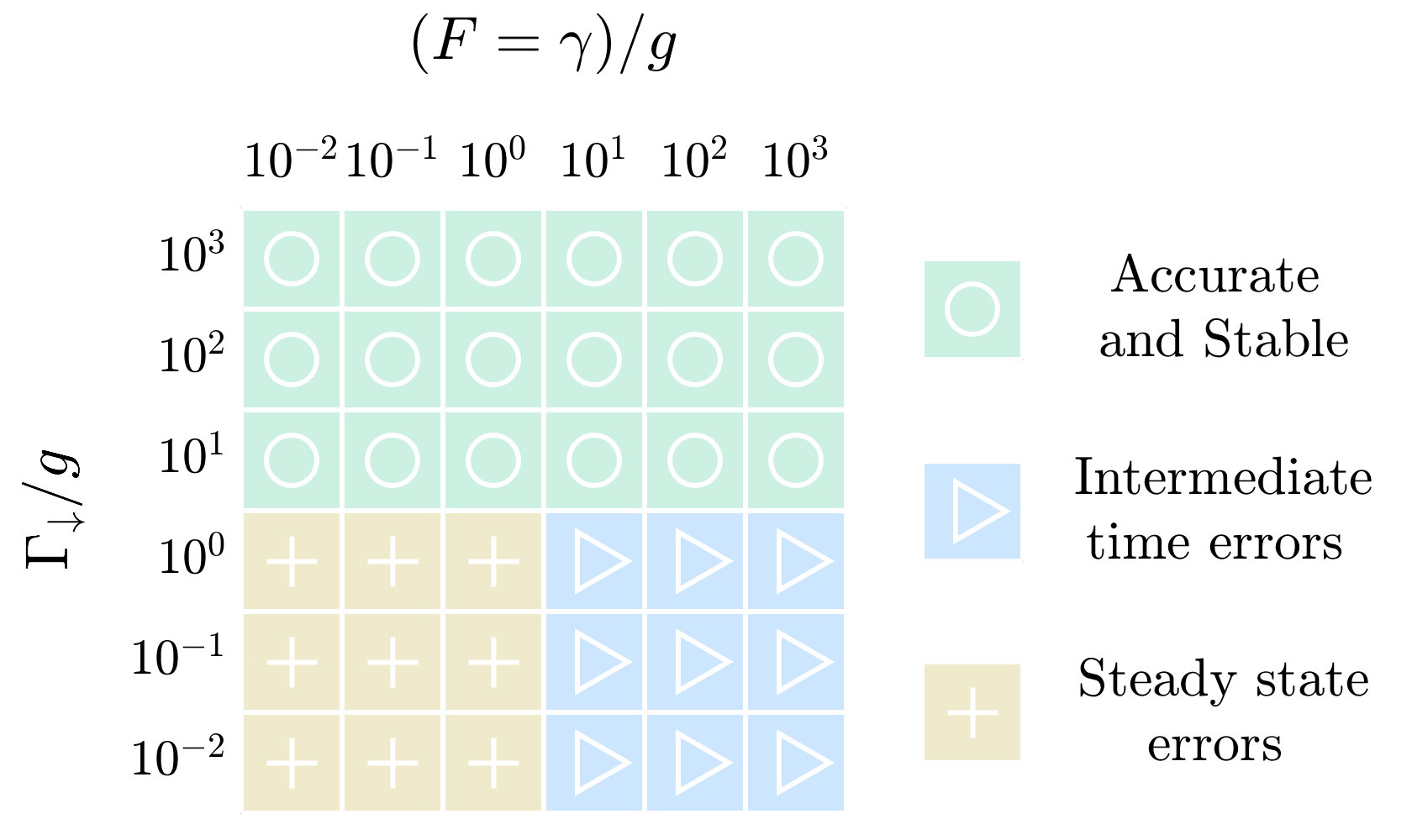}{{\bf Stability diagram for the SU(2) coherent state kernel.} Simulation stability over a broad range of $\gamma$ and $\Gamma_\downarrow$. Green squares indicate accurate, stable simulations throughout; blue squares show transient intermediate-time \mbox{errors} but recover the correct steady state; yellow squares indicate errors that persist into the steady state. \label{StabSU2}}{b}{1}

To computationally handle the issue of the spin up state being at $z_j,w_j\to\infty$ in this formalism, we use a \mbox{spin-dependent} mapping \cite{Ng13}, where our simulation \mbox{variables} $\{\bar{z}_j,\bar{w}_j\}$ produced during the numerical integration dynamically switch between the above mapping which has $z,w=0$ in the down spin, and the opposite construction with $z,w=0$ in the opposite pole of the Bloch sphere as the trajectories evolve: \pagebreak
\eqy{SdM}{
\bar{z}_j =&\,\begin{cases} z_j\, , & |z_j| \leq 1 \\ 1/z_j^*\, , & |z_j| > 1 \end{cases}\, , \nonu\\
\bar{w}_j =&\,\begin{cases} w_j\, , & |w_j| \leq 1 \\ 1/w_j^*\, , & |w_j| > 1\end{cases}\, .
}
In terms of the numerical integration algorithm, this is done by checking after each time step of the integration, and remapping the variables $\bar{z}_j\to 1/\bar{z}_j^*$, $\bar{w}_j\to 1/\bar{w}_j^*$ \mbox{respectively} when they are outside the unit circle in the complex plane; a flag for each $\bar{z}_j$ and $\bar{w}_j$ that can take \mbox{values} $\pm1$ keeps track of whether that variable is \mbox{representing} $z_j$ or $1/z_j^*$ (and equivalently for $w_j$) at that time step.  It is then used to determine which of the set of stochastic differential equations for $\bar{z}_j$ and $\bar{w}_j$, formed by applying the cases \eqn{SdM} to \eqn{ItoSU2}, to evolve with for the \mbox{upcoming} time step, as well as how to calculate \mbox{observables} from the output for that time.  

\figv{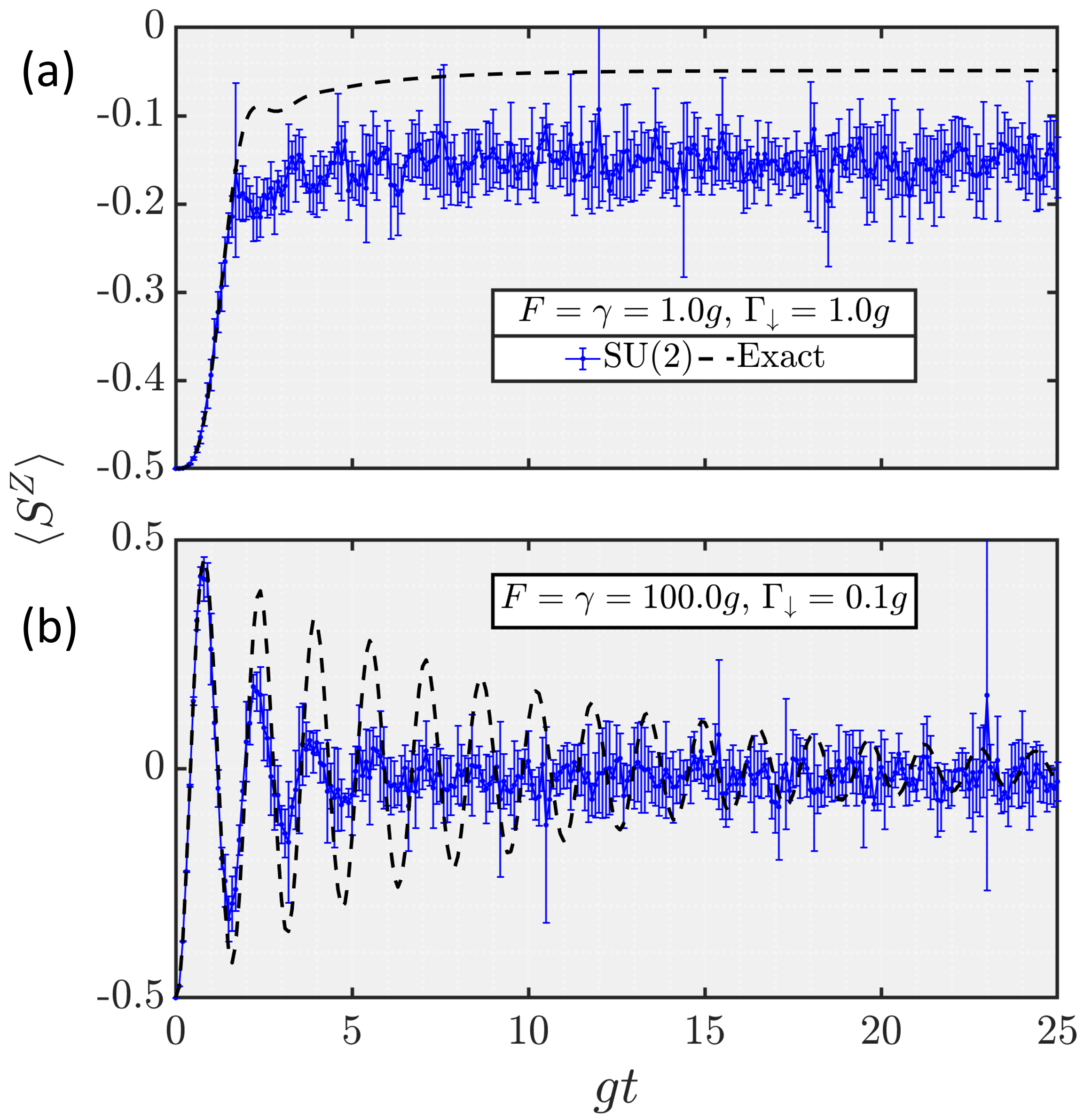}{{\bf Examples of systematic errors in SU(2) \mbox{kernel} simulations.} a) $\expect{\op{S}^Z}$ for $F=\gamma=\Gamma_\downarrow=g$, where errors emerge at finite times and persist into the steady state (yellow squares in Fig.~\ref{StabSU2}). b) $\expect{\op{S}^Z}$ for $F=\gamma=100g$, $\Gamma_\downarrow=0.1g$, where errors appear as overdamped oscillations at intermediate times but converge toward the correct steady state (blue squares in Fig.~\ref{StabSU2}). \label{SU2error}}{b}{1}

\vspace{-3mm}
In Fig.~\ref{StabSU2}, we summarise the equivalent stability \mbox{analysis} to Fig.~\ref{StabGDM} using this method.  The stable \mbox{regions} follow the same pattern here as for the GDM method, \mbox{except} that rather than a clear instability marking the cases where the method is not valid, it experiences \mbox{instead} one of two kinds of systematic errors compared to the exact results when $\Gamma_\downarrow \leq g$.  For $\gamma \leq g$ \mbox{systematic} errors manifest after a finite time and remain into the steady state (see example in Fig.~\ref{SU2error}(a)).  In the cases with $\gamma > g$, systematic errors occur at intermediate times, \mbox{often} manifesting as an over-damping of \mbox{oscillations} compared to the exact result, but seemingly the \mbox{correct} steady state is reached, at least within the \mbox{numerical} \mbox{error} bars, for $n$ and $\expect{\op{S}^Z}$ (see examples in Fig.~\ref{SU2error}(b) and Fig.~\ref{xKernComp}).  While the latter case may still be useful for observing steady state properties, one should be cautious in assuming that the \mbox{distribution} reached by the SU(2) kernel simulations in the steady state will give the \mbox{correct} values for all \mbox{observables,} due to the errors present at \mbox{earlier} times.

\subsection{Stratonovich corrections}
\label{StrC}

The stochastic differential equations \eqn{Ito_full} and \eqn{ItoSU2} are written in the form corresponding to the Ito \mbox{stochastic} calculus.  When numerically integrating the equations, only the simple Euler method produces the Ito \mbox{integral;} other more sophisticated integration algorithms most commonly calculate the Stratonovich integral as they take advantage of the rules of deterministic calculus \mbox{remaining} intact in the Stratonovich calculus \mbox{\cite{StochMech,DeuarPhD,XMDS_CPhysComm.184.201,Drummond91} }
(though other corrections are also possible depending on the time stepping algorithm \cite{tcorr}, in xmds2 algorithms are implemented so that stochastic equations should be entered in Stratonovich form \cite{XMDS_CPhysComm.184.201}).  Converting from the Ito calculus to the Stratonovich calculus leads to \mbox{possible} corrections to the drift which depend on the noise matrix:
\eq{StrCeq}{
A_m \to A_m - \frac{1}{2} B_{k,l}\frac{\partial B_{m,l}}{\partial v_k} \, ,
}
where summation over repeated indices is again implied.  

For our formulation of the SU(2) kernel \mbox{stochastic} equations in section~\ref{SESU2} above, $B_g$ and $B_\downarrow$ give zero \mbox{corrections,} while $B_\phi$ gives corrections of 
\eq{SCSU2phi}{
\dada{z_j}{t} \to \dada{z_j}{t} + \frac{\Gamma_\phi z_j}{2}\, , \quad \dada{w_j}{t} \to \dada{w_j}{t} + \frac{\Gamma_\phi w_j}{2}\, ,
}
which exactly cancel the Ito drift contribution from $\Gamma_\phi$ terms, leading to the Stratonovich stochastic differential equations:
\eqy{StratSU2}{
\dada{z_j}{t} =&\, (-i\omega_0+i\sqrt{\Gamma_\phi}\xi_{\phi,j}-\frac{\Gamma_\downarrow}{2}\frac{1+3z_j w_j}{1+z_j w_j})z_j \nonu 
\\& +\sqrt{\Gamma_\downarrow}z_j^2\eta_{\downarrow,j} -i\frac{h}{2}(1-z_j^2) \nonu 
\\& -ig(\alpha_j-\beta_j z_j^2) +z_j^2\sqrt{ig}\eta_{Z,j} \, , \nonu\\
\dada{w_j}{t} =&\, (i\omega_0-i\sqrt{\Gamma_\phi}\xi_{\phi,j}-\frac{\Gamma_\downarrow}{2}\frac{1+3z_j w_j}{1+z_j w_j})w_j \nonu 
\\& +\sqrt{\Gamma_\downarrow}w_j^2\eta_{\downarrow,j}^* +i\frac{h}{2}(1-w_j^2) \nonu 
\\& +ig(\beta_j-\alpha_j w_j^2) -w_j^2\sqrt{ig}\eta_{W,j} \, , \nonu\\
\dada{\alpha_j}{t} =&\, (-i\omega_C-\frac{\gamma}{2})\alpha_j -ig\frac{z_j}{1+z_jw_j} \nonu 
\\& -iF +i\sum_{j'}J_{j,j'}\alpha_{j'} +\sqrt{ig}\eta^*_{Z,j}\, , \nonu\\
\dada{\beta_j}{t} =&\, (i\omega_C-\frac{\gamma}{2})\beta_j +ig\frac{w_j}{1+z_jw_j} \nonu 
\\& +iF -i\sum_{j'}J_{j,j'}\beta_{j'} +\sqrt{ig}\eta^*_{W,j}\, ,
}

Notably here, the spin dephasing $\Gamma_\phi$ now produces only noise contributions, which essentially act like a random variation of the spin energy $\omega_0 \to \omega_0-\sqrt{\Gamma_\phi}\xi_{\phi,j}$; this was similarly true for the Schwinger Boson approach equations \eqn{SchwingStrat}.  In contrast, for the GDM kernel equations dephasing instead contributes no noise and only a decay of $\{z_j,w_j\}$ in their respective drifts.  

For the GDM kernel, the noise matrix $B$ in equation \eqn{diffG} gives a Stratonovich correction
\eq{SCxkern}{
\dada{\alpha_j}{t} \to \dada{\alpha_j}{t} - \frac{ig z_j}{2}\, , \quad \dada{\beta_j}{t} \to \dada{\beta_j}{t} + \frac{ig w_j}{2}\, ,
}
for any constant diffusion gauge $G$.  This leads to the Stratonovich form of the stochastic differential equations in \eqn{Ito_full} as
\eqa{Strat_full}{
\dada{z_j}{t} =&\, (-i\omega_0-\frac{\Gamma_\phi}{2}-\frac{\Gamma_\downarrow}{2})z_j -i(\frac{h}{2}+g\alpha_j)(1-2x_j) \nonu 
\\& +\sqrt{\frac{ig}{G}}(z_j\eta_{Z,j} +z_j\eta_{W,j} +\eta_{B,j})\, , \nonu\\
\dada{w_j}{t} =&\, (i\omega_0-\frac{\Gamma_\phi}{2}-\frac{\Gamma_\downarrow}{2})w_j +i(\frac{h}{2}+g\beta_j)(1-2x_j) \nonu 
\\& +\sqrt{\frac{ig}{G}}(w_j\eta_{Z,j} +w_j\eta_{W,j} +\eta_{A,j})\, , \nonu\\
\dada{x_j}{t} =&\, \frac{ih}{2}(z_j-w_j)+ig(z_j\beta_j-w_j\alpha_j) -\Gamma_\downarrow x_j \nonu 
\\& +\sqrt{\frac{ig}{G}}x_j(\eta_{Z,j} +\eta_{W,j})\, , \nonu\\
\dada{\alpha_j}{t} =&\, (-i\omega_C-\frac{\gamma}{2})\alpha_j -\frac{3igz_j}{2} -iF +i\sum_{j'}J_{j,j'}\alpha_{j'} \nonu 
\\& +\sqrt{igG}(z_j\eta^*_{Z,j} -x_j\eta^*_{A,j})\, , \nonu\\
\dada{\beta_j}{t} =&\, (i\omega_C-\frac{\gamma}{2})\beta_j +\frac{3igw_j}{2} +iF -i\sum_{j'}J_{j,j'}\beta_{j'} \nonu 
\\& -\sqrt{igG}(w_j\eta^*_{W,j} -x_j\eta^*_{B,j})\, .  
}
If the diffusion gauge $G(\vec{v})$ is instead allowed to have an arbitrary dependence on the phase space variables, then due to the derivatives in the definition \eqn{StrCeq} this may alter the Stratonovich correction significantly depending on the choice.  


\section*{Code Availability}

Codes for the GDM kernel simulations are available on Github at \url{https://github.com/A-Ferrier/SBPP}.  Code for other methods discussed can be made available from the corresponding author upon reasonable request.

\section*{Data Availability}

The data generated and analysed in this work can be made available from the corresponding author upon \mbox{reasonable} request.

\acknowledgments

Some 3D renders used in the construction of Fig.~\ref{DDJC_sys} were originally provided courtesy of Connor McKeever.  This research was partly carried out using the high performance computing cluster in the London Centre for Nanotechnology.  The authors acknowledge the use of the UCL Myriad High Performance Computing Facility (Myriad@UCL), and associated support services, in the completion of this work.  A.F. and M.H.S. acknowledge financial support from EPSRC (Grant No.~EP/V026496/1) and from the EUCENTRAL project, funded by the European Union under Horizon Europe (project No.~101186579).  P.D. acknowledges support by the National Science Centre (Poland) grant No.~2018/31/B/ST2/01871. Y.S. acknowledges support from EPSRC (Grant No.~EP/S021582/1).

\section*{Author Contributions}

A.F. and P.D. developed the theoretical formalism and derived the stochastic differential equations.  A.F. ran the stochastic simulations and wrote the initial draft of the manuscript.  Y.S. ran the direct solutions of the master equation used for benchmarking.  P.D. and M.H.S. provided supervision and guidance for the direction of the research.  A.F., P.D. and M.H.S. contributed revisions to the manuscript.  All authors read and approved the final manuscript.

\section*{Competing Interests}

The authors declare no competing interests.  

\bibliography{Bibliographies/Books,Bibliographies/Miscellarticles,Bibliographies/PhysRevLett,Bibliographies/PhysRevB,Bibliographies/ArXiv,Bibliographies/OtherPhysRev,Bibliographies/Nature_all,Bibliographies/artnew}

@misc{dunham2025,
      title={{Efficient Time Evolution of 2D Open-Quantum Lattice Models with Long-Range Interactions using Tensor Networks}}, 
      author={Jack Dunham and Marzena H. Szymańska},
      year={2025},
      eprint={2512.01781},
      archivePrefix={arXiv},
      primaryClass={quant-ph},
      url={https://arxiv.org/abs/2512.01781}, 
}

@misc{Yi_arXiv,
      title={{Stochastic phase-space simulation of multimode cat states via the positive-P representation}}, 
      author={Yi Shi and Alex Ferrier and Piotr Deuar and Eran Ginossar and Marzena Szymanska},
      year={2026},
      eprint={2601.07049},
      archivePrefix={arXiv},
      primaryClass={quant-ph},
      url={https://arxiv.org/abs/2601.07049}, 
}

@misc{cao2026neuralgaugeprepresentationopen,
      title={{Neural Gauge-P Representation for Open Quantum Dynamics of Interacting Bosons}}, 
      author={Xiaodong Cao and Zhicheng Zhong},
      year={2026},
      eprint={2607.17534},
      archivePrefix={arXiv},
      primaryClass={quant-ph},
      url={https://arxiv.org/abs/2607.17534}, 
}

@article{a,title="UNKNOWN",author="UNKNOWN",journal="UNKNOWN",year = "UNKNOWN"}

@article{XMDS_CPhysComm.184.201, title = "{XMDS2: Fast, scalable simulation of coupled stochastic partial differential equations}", journal = "Computer Physics Communications", volume = "184", number = "1", pages = "201 - 208", year = "2013", issn = "0010-4655", doi = "https://doi.org/10.1016/j.cpc.2012.08.016", author = "Graham R. Dennis and Joseph J. Hope and Mattias T. Johnsson"}

@article{RevModPhys.93.025005,
  title = {Circuit quantum electrodynamics},
  author = {Blais, Alexandre and Grimsmo, Arne L. and Girvin, S. M. and Wallraff, Andreas},
  journal = {Rev. Mod. Phys.},
  volume = {93},
  issue = {2},
  pages = {025005},
  numpages = {72},
  year = {2021},
  month = {May},
  publisher = {American Physical Society},
  doi = {10.1103/RevModPhys.93.025005},
  url = {https://link.aps.org/doi/10.1103/RevModPhys.93.025005}
}

@article{PhysRevX.5.011022,
  title = {{Many-Body Quantum Spin Dynamics with Monte Carlo Trajectories on a Discrete Phase Space}},
  author = {Schachenmayer, J. and Pikovski, A. and Rey, A. M.},
  journal = {Phys. Rev. X},
  volume = {5},
  issue = {1},
  pages = {011022},
  numpages = {10},
  year = {2015},
  month = {Feb},
  publisher = {American Physical Society},
  doi = {10.1103/PhysRevX.5.011022}
}

@article{PRXQuantum.2.010319,
  title = {Fully Quantum Scalable Description of Driven-Dissipative Lattice Models},
  author = {Deuar, Piotr and Ferrier, Alex and Matuszewski, Micha\l{} and Orso, Giuliano and Szyma\ifmmode \acute{n}\else \'{n}\fi{}ska, Marzena H.},
  journal = {PRX Quantum},
  volume = {2},
  issue = {1},
  pages = {010319},
  numpages = {19},
  year = {2021},
  month = {Feb},
  publisher = {American Physical Society},
  doi = {10.1103/PRXQuantum.2.010319}
}

@article{PhysRevX.11.021035,
  title = {Stable {iPEPO} Tensor-Network Algorithm for Dynamics of Two-Dimensional Open Quantum Lattice Models},
  author = {Mc Keever, C. and Szyma\ifmmode \acute{n}\else \'{n}\fi{}ska, M. H.},
  journal = {Phys. Rev. X},
  volume = {11},
  issue = {2},
  pages = {021035},
  numpages = {17},
  year = {2021},
  month = {May},
  publisher = {American Physical Society},
  doi = {10.1103/PhysRevX.11.021035},
  url = {https://link.aps.org/doi/10.1103/PhysRevX.11.021035}
}

@article{PhysRevA.105.013716,
  title = {Realistic simulations of spin squeezing and cooperative coupling effects in large ensembles of interacting two-level systems},
  author = {Huber, Julian and Rey, Ana Maria and Rabl, Peter},
  journal = {Phys. Rev. A},
  volume = {105},
  issue = {1},
  pages = {013716},
  numpages = {14},
  year = {2022},
  month = {Jan},
  publisher = {American Physical Society},
  doi = {10.1103/PhysRevA.105.013716},
  url = {https://link.aps.org/doi/10.1103/PhysRevA.105.013716}
}

@article{PhysRevResearch.5.043288,
  title = {{Trajectory-resolved Weiss fields for quantum spin dynamics}},
  author = {Begg, S. E. and Green, A. G. and Bhaseen, M. J.},
  journal = {Phys. Rev. Res.},
  volume = {5},
  issue = {4},
  pages = {043288},
  numpages = {13},
  year = {2023},
  month = {Dec},
  publisher = {American Physical Society},
  doi = {10.1103/PhysRevResearch.5.043288},
  url = {https://link.aps.org/doi/10.1103/PhysRevResearch.5.043288}
}

@article{PRXQuantum.4.030304,
  title = {Quantum and Classical Correlations in Open Quantum Spin Lattices via Truncated-Cumulant Trajectories},
  author = {Verstraelen, Wouter and Huybrechts, Dolf and Roscilde, Tommaso and Wouters, Michiel},
  journal = {PRX Quantum},
  volume = {4},
  issue = {3},
  pages = {030304},
  numpages = {28},
  year = {2023},
  month = {Jul},
  publisher = {American Physical Society},
  doi = {10.1103/PRXQuantum.4.030304},
  url = {https://link.aps.org/doi/10.1103/PRXQuantum.4.030304}
}

@article{PhysRevX.15.021073,
  title = {Hybrid Quantum-Classical Stochastic Approach to Dissipative Spin-Boson Models},
  author = {Kamar, Naushad A. and Maghrebi, Mohammad},
  journal = {Phys. Rev. X},
  volume = {15},
  issue = {2},
  pages = {021073},
  numpages = {31},
  year = {2025},
  month = {May},
  publisher = {American Physical Society},
  doi = {10.1103/PhysRevX.15.021073},
  url = {https://link.aps.org/doi/10.1103/PhysRevX.15.021073}
}

@article{PhysRevB.104.024408,
  title = {{Time-evolving Weiss fields in the stochastic approach to quantum spins}},
  author = {Begg, S. E. and Green, A. G. and Bhaseen, M. J.},
  journal = {Phys. Rev. B},
  volume = {104},
  issue = {2},
  pages = {024408},
  numpages = {9},
  year = {2021},
  month = {Jul},
  publisher = {American Physical Society},
  doi = {10.1103/PhysRevB.104.024408},
  url = {https://link.aps.org/doi/10.1103/PhysRevB.104.024408}
}

@article{PhysRevLett.122.250501,
  title = {{Variational Quantum Monte Carlo Method with a Neural-Network Ansatz for Open Quantum Systems}},
  author = {Nagy, Alexandra and Savona, Vincenzo},
  journal = {Phys. Rev. Lett.},
  volume = {122},
  issue = {25},
  pages = {250501},
  numpages = {6},
  year = {2019},
  month = {Jun},
  publisher = {American Physical Society},
  doi = {10.1103/PhysRevLett.122.250501},
  url = {https://link.aps.org/doi/10.1103/PhysRevLett.122.250501}
}

@article{PhysRevLett.122.250502,
  title = {Neural-Network Approach to Dissipative Quantum Many-Body Dynamics},
  author = {Hartmann, Michael J. and Carleo, Giuseppe},
  journal = {Phys. Rev. Lett.},
  volume = {122},
  issue = {25},
  pages = {250502},
  numpages = {6},
  year = {2019},
  month = {Jun},
  publisher = {American Physical Society},
  doi = {10.1103/PhysRevLett.122.250502},
  url = {https://link.aps.org/doi/10.1103/PhysRevLett.122.250502}
}

@article{PhysRevLett.122.250503,
  title = {Variational Neural-Network Ansatz for Steady States in Open Quantum Systems},
  author = {Vicentini, Filippo and Biella, Alberto and Regnault, Nicolas and Ciuti, Cristiano},
  journal = {Phys. Rev. Lett.},
  volume = {122},
  issue = {25},
  pages = {250503},
  numpages = {6},
  year = {2019},
  month = {Jun},
  publisher = {American Physical Society},
  doi = {10.1103/PhysRevLett.122.250503},
  url = {https://link.aps.org/doi/10.1103/PhysRevLett.122.250503}
}

@article{Olsen09,
title = "{Numerical representation of quantum states in the positive-P and Wigner representations}",
journal = "Optics Communications ",
volume = "282",
number = "19",
pages = "3924 - 3929",
year = "2009",
note = "",
issn = "0030-4018",
doi = "http://dx.doi.org/10.1016/j.optcom.2009.06.033",
url = "http://www.sciencedirect.com/science/article/pii/S0030401809005963",
author = "M.K. Olsen and A.S. Bradley"
}

@article{Deuar07,
  title = {{Correlations in a BEC Collision: First-Principles Quantum Dynamics with 150\,000 Atoms}},
  author = {Deuar, P. and Drummond, P. D.},
  journal = {Phys. Rev. Lett.},
  volume = {98},
  issue = {12},
  pages = {120402},
  numpages = {4},
  year = {2007},
  month = {Mar},
  publisher = {American Physical Society},
  doi = {10.1103/PhysRevLett.98.120402},
  url = {http://link.aps.org/doi/10.1103/PhysRevLett.98.120402}
}

@article{Deuar09b,
  title = {Simulation of Complete Many-Body Quantum Dynamics Using Controlled Quantum-Semiclassical Hybrids},
  author = {Deuar, P.},
  journal = {Phys. Rev. Lett.},
  volume = {103},
  issue = {13},
  pages = {130402},
  numpages = {4},
  year = {2009},
  month = {Sep},
  publisher = {American Physical Society},
  doi = {10.1103/PhysRevLett.103.130402},
  url = {http://link.aps.org/doi/10.1103/PhysRevLett.103.130402}
}

@article{Deuar06a,
  author={P Deuar and P D Drummond},
  title={First-principles quantum dynamics in interacting {Bose} gases: I. The positive P representation},
  journal={Journal of Physics A: Mathematical and General},
  volume={39},
  number={5},
  pages={1163},
  url={http://stacks.iop.org/0305-4470/39/i=5/a=010},
  doi={10.1088/0305-4470/39/5/010},
  year={2006}
}

@article{Deuar06b,
  author={P Deuar and P D Drummond},
  title={First-principles quantum dynamics in interacting {Bose} gases: {II}: stochastic gauges},
  journal={Journal of Physics A: Mathematical and General},
  volume={39},
  number={5},
  pages={2723},
  doi={10.1088/0305-4470/39/11/011},
  year={2006}
}

@article{Lewis-Swan14,
  title = {Proposal for demonstrating the {Hong–Ou–Mandel} effect with matter waves},
  author = {Lewis-Swan, R. J. and Kheruntsyan, K. V.},
  journal = {Nature Commun.},
  volume = {5},
  pages = {3752},
  year = {2014},
  month = {Apr},
  doi = {10.1038/ncomms4752}
}

@article{Kheruntsyan12,
  title = {{Violation of the Cauchy-Schwarz Inequality with Matter Waves}},
  author = {Kheruntsyan, K. V. and Jaskula, J.-C. and Deuar, P. and Bonneau, M. and Partridge, G. B. and Ruaudel, J. and Lopes, R. and Boiron, D. and Westbrook, C. I.},
  journal = {Phys. Rev. Lett.},
  volume = {108},
  issue = {26},
  pages = {260401},
  numpages = {5},
  year = {2012},
  month = {Jun},
  publisher = {American Physical Society},
  doi = {10.1103/PhysRevLett.108.260401},
  url = {http://link.aps.org/doi/10.1103/PhysRevLett.108.260401}
}

@article{Drummond80,
  author={P D Drummond and C W Gardiner},
  title={{Generalised P-representations in quantum optics}},
  journal={Journal of Physics A: Mathematical and General},
  volume={13},
  number={7},
  pages={2353},
  url={http://stacks.iop.org/0305-4470/13/i=7/a=018},
  year={1980}
}

@article{Martin10b,
  title = {Quantum and Thermal Effects of Dark Solitons in a One-Dimensional {Bose} Gas},
  author = {Martin, A. D. and Ruostekoski, J.},
  journal = {Phys. Rev. Lett.},
  volume = {104},
  issue = {19},
  pages = {194102},
  numpages = {4},
  year = {2010},
  month = {May},
  publisher = {American Physical Society},
  doi = {10.1103/PhysRevLett.104.194102},
  url = {http://link.aps.org/doi/10.1103/PhysRevLett.104.194102}
}

@article{Sinatra02,
  author={Alice Sinatra and Carlos Lobo and Yvan Castin},
  title={{The truncated Wigner method for {Bose}-condensed gases: limits of validity and applications}},
  journal={Journal of Physics B: Atomic, Molecular and Optical Physics},
  volume={35},
  number={17},
  pages={3599},
  url={http://stacks.iop.org/0953-4075/35/i=17/a=301},
  year={2002}
}

@article{Steel98,
  title = {Dynamical quantum noise in trapped {Bose}-{Einstein} condensates},
  author = {Steel, M. J. and Olsen, M. K. and Plimak, L. I. and Drummond, P. D. and Tan, S. M. and Collett, M. J. and Walls, D. F. and Graham, R.},
  journal = {Phys. Rev. A},
  volume = {58},
  issue = {6},
  pages = {4824--4835},
  numpages = {0},
  year = {1998},
  month = {Dec},
  publisher = {American Physical Society},
  doi = {10.1103/PhysRevA.58.4824},
  url = {http://link.aps.org/doi/10.1103/PhysRevA.58.4824}
}

@article{Norrie05,
  title = {Quantum Turbulence in Condensate Collisions: An Application of the Classical Field Method},
  author = {Norrie, A. A. and Ballagh, R. J. and Gardiner, C. W.},
  journal = {Phys. Rev. Lett.},
  volume = {94},
  issue = {4},
  pages = {040401},
  numpages = {4},
  year = {2005},
  month = {Jan},
  publisher = {American Physical Society},
  doi = {10.1103/PhysRevLett.94.040401},
  url = {http://link.aps.org/doi/10.1103/PhysRevLett.94.040401}
}

@article{Hoffmann08,
  title = {Hybrid phase-space simulation method for interacting {Bose} fields},
  author = {Hoffmann, Scott E. and Corney, Joel F. and Drummond, Peter D.},
  journal = {Phys. Rev. A},
  volume = {78},
  issue = {1},
  pages = {013622},
  numpages = {12},
  year = {2008},
  month = {Jul},
  publisher = {American Physical Society},
  doi = {10.1103/PhysRevA.78.013622},
  url = {http://link.aps.org/doi/10.1103/PhysRevA.78.013622}
}

@article{Carter87,
  title = {Squeezing of quantum solitons},
  author = {Carter, S. J. and Drummond, P. D. and Reid, M. D. and Shelby, R. M.},
  journal = {Phys. Rev. Lett.},
  volume = {58},
  issue = {18},
  pages = {1841--1844},
  numpages = {0},
  year = {1987},
  month = {May},
  publisher = {American Physical Society},
  doi = {10.1103/PhysRevLett.58.1841},
  url = {http://link.aps.org/doi/10.1103/PhysRevLett.58.1841}
}

@phdthesis{DeuarPhD,
   author = {P. Deuar},
   title = {First-principles quantum simulations of many-mode open interacting {Bose} gases using stochastic gauge methods},
   school = {University of Queensland, arXiv:cond-mat/0507023},
   month = {February},
   year = {2005}
}

@book{StochMech,
  author={C. W. Gardiner},
  title={Stochastic Methods}, 
  publisher = {Springer-Verlag},
  edition = {4th},
  address = {Berlin, Heidelberg},
  year = {2009}
}

@article{Gilchrist97,
  title = {{Positive P representation: Application and validity}},
  author = {Gilchrist, A. and Gardiner, C. W. and Drummond, P. D.},
  journal = {Phys. Rev. A},
  volume = {55},
  issue = {4},
  pages = {3014--3032},
  numpages = {0},
  year = {1997},
  month = {Apr},
  publisher = {American Physical Society},
  doi = {10.1103/PhysRevA.55.3014},
  url = {http://link.aps.org/doi/10.1103/PhysRevA.55.3014}
}

@article{Wouters09,
  title = {Stochastic classical field model for polariton condensates},
  author = {Wouters, Michiel and Savona, Vincenzo},
  journal = {Phys. Rev. B},
  volume = {79},
  issue = {16},
  pages = {165302},
  numpages = {10},
  year = {2009},
  month = {Apr},
  publisher = {American Physical Society},
  doi = {10.1103/PhysRevB.79.165302},
  url = {http://link.aps.org/doi/10.1103/PhysRevB.79.165302}
}

@article{Drummond91,
title = "Computer simulations of multiplicative stochastic differential equations",
journal = "Journal of Computational Physics",
volume = "93",
number = "1",
pages = "144 - 170",
year = "1991",
note = "",
issn = "0021-9991",
author = "P.D Drummond and I.K Mortimer"
}

@article{Schmidt13,
author = {Schmidt, Sebastian and Koch, Jens},
title = {{Circuit QED lattices: Towards quantum simulation with superconducting circuits}},
journal = {Annalen der Physik},
volume = {525},
number = {6},
pages = {395-412},
doi = {10.1002/andp.201200261},
url = {https://onlinelibrary.wiley.com/doi/abs/10.1002/andp.201200261},
year = {2013}
}

@article{Biondi15,
  title = {Incompressible Polaritons in a Flat Band},
  author = {Biondi, Matteo and van Nieuwenburg, Evert P. L. and Blatter, Gianni and Huber, Sebastian D. and Schmidt, Sebastian},
  journal = {Phys. Rev. Lett.},
  volume = {115},
  issue = {14},
  pages = {143601},
  numpages = {6},
  year = {2015},
  month = {Sep},
  publisher = {American Physical Society},
  doi = {10.1103/PhysRevLett.115.143601},
  url = {https://link.aps.org/doi/10.1103/PhysRevLett.115.143601}
}

@article{Deuar02,
  title = {{Gauge $P$ representations for quantum-dynamical problems: Removal of boundary terms}},
  author = {Deuar, P. and Drummond, P. D.},
  journal = {Phys. Rev. A},
  volume = {66},
  issue = {3},
  pages = {033812},
  numpages = {16},
  year = {2002},
  month = {Sep},
  publisher = {American Physical Society},
  doi = {10.1103/PhysRevA.66.033812},
  url = {https://link.aps.org/doi/10.1103/PhysRevA.66.033812}
}

@article{Fink17,
  title = {Observation of the Photon-Blockade Breakdown Phase Transition},
  author = {Fink, J. M. and Dombi, A. and Vukics, A. and Wallraff, A. and Domokos, P.},
  journal = {Phys. Rev. X},
  volume = {7},
  issue = {1},
  pages = {011012},
  numpages = {9},
  year = {2017},
  month = {Jan},
  publisher = {American Physical Society},
  doi = {10.1103/PhysRevX.7.011012},
  url = {https://link.aps.org/doi/10.1103/PhysRevX.7.011012}
}

@article{Fitzpatrick17,
  title = {{Observation of a Dissipative Phase Transition in a One-Dimensional Circuit QED Lattice}},
  author = {Fitzpatrick, Mattias and Sundaresan, Neereja M. and Li, Andy C. Y. and Koch, Jens and Houck, Andrew A.},
  journal = {Phys. Rev. X},
  volume = {7},
  issue = {1},
  pages = {011016},
  numpages = {8},
  year = {2017},
  month = {Feb},
  publisher = {American Physical Society},
  doi = {10.1103/PhysRevX.7.011016},
  url = {https://link.aps.org/doi/10.1103/PhysRevX.7.011016}
}

@article{Houck12,
author = {Houck, A. A.  and Türeci, H. E. and Koch, J.},
year = {2012},
title = {On-chip quantum simulation with superconducting circuits},
journal = {Nature Physics},
pages = {292--299},
volume = {8},
issue = {4},
doi = {10.1038/nphys2251}
}

@unpublished{ppdd,
  author = {Deuar, P. and Ferrier, A. and Matuszewski, M. and Orso, G. and  Szyma\'nska, M. H. },
  title = {Fully quantum scalable description of driven dissipative lattice models},
  note = {In preparation},
  year = {2020}
}

@article{Ng13,
  title = {Simulation of the dynamics of many-body quantum spin systems using phase-space techniques},
  author = {Ng, Ray and S\o{}rensen, Erik S. and Deuar, Piotr},
  journal = {Phys. Rev. B},
  volume = {88},
  issue = {14},
  pages = {144304},
  numpages = {14},
  year = {2013},
  month = {Oct},
  publisher = {American Physical Society},
  doi = {10.1103/PhysRevB.88.144304},
  url = {https://link.aps.org/doi/10.1103/PhysRevB.88.144304}
}

@article{Mandt15,
	doi = {10.1088/1367-2630/17/5/053018},
	url = {https://doi.org/10.1088%2F1367-2630%2F17%2F5%2F053018},
	year = 2015,
	month = {may},
	publisher = {{IOP} Publishing},
	volume = {17},
	number = {5},
	pages = {053018},
	author = {Stephan Mandt and Darius Sadri and Andrew A Houck and Hakan E Türeci},
	title = {Stochastic differential equations for quantum dynamics of spin-boson networks},
	journal = {New Journal of Physics}
}

@article{Barry08,
  title = {{Qubit phase space: SU$(n)$ coherent-state $P$ representations}},
  author = {Barry, D. W. and Drummond, P. D.},
  journal = {Phys. Rev. A},
  volume = {78},
  issue = {5},
  pages = {052108},
  numpages = {13},
  year = {2008}
}

@article{tcorr,
  doi = {https://doi.org/10.22331/q-2021-05-10-455},
  _url = {https://doi.org/10.22331/q-2021-05-10-455},
  title = {Multi-time correlations in the positive-{P}, {Q}, and doubled phase-space representations},
  author = {Deuar, Piotr},
  journal = {{Quantum}},
  _issn = {2521-327X},
  publisher = {{Verein zur F{\"{o}}rderung des Open Access Publizierens in den Quantenwissenschaften}},
  volume = {5},
  pages = {455},
  _month = may,
  year = {2021}
}

@article{Carusotto05,
  title = {Spontaneous microcavity-polariton coherence across the parametric threshold: Quantum Monte Carlo studies},
  author = {Carusotto, Iacopo and Ciuti, Cristiano},
  journal = {Phys. Rev. B},
  volume = {72},
  issue = {12},
  pages = {125335},
  numpages = {13},
  year = {2005},
  month = {Sep},
  publisher = {American Physical Society},
  doi = {10.1103/PhysRevB.72.125335},
  url = {https://link.aps.org/doi/10.1103/PhysRevB.72.125335}
}

@article{Ballarini20,
   title={Directional Goldstone waves in polariton condensates close to equilibrium},
   volume={11},
   ISSN={2041-1723},
   url={http://dx.doi.org/10.1038/s41467-019-13733-x},
   doi={10.1038/s41467-019-13733-x},
   number={1},
   pages={217},
   journal={Nature Communications},
   publisher={Springer Science and Business Media LLC},
   author={Ballarini, Dario and Caputo, Davide and Dagvadorj, Galbadrakh and Juggins, Richard and Giorgi, Milena De and Dominici, Lorenzo and West, Kenneth and Pfeiffer, Loren N. and Gigli, Giuseppe and Szymańska, Marzena H. and et al.},
   year={2020},
   month={Jan}
}

@article{Comaron18,
  title = {Dynamical Critical Exponents in Driven-Dissipative Quantum Systems},
  author = {Comaron, P. and Dagvadorj, G. and Zamora, A. and Carusotto, I. and Proukakis, N. P. and Szyma\ifmmode \acute{n}\else \'{n}\fi{}ska, M. H.},
  journal = {Phys. Rev. Lett.},
  volume = {121},
  issue = {9},
  pages = {095302},
  numpages = {6},
  year = {2018},
  month = {Aug},
  publisher = {American Physical Society},
  doi = {10.1103/PhysRevLett.121.095302},
  url = {https://link.aps.org/doi/10.1103/PhysRevLett.121.095302}
}

@article{Caputo17,
   title={Topological order and thermal equilibrium in polariton condensates},
   volume={17},
   ISSN={1476-4660},
   url={http://dx.doi.org/10.1038/nmat5039},
   doi={10.1038/nmat5039},
   number={2},
   journal={Nature Materials},
   publisher={Springer Science and Business Media LLC},
   author={Caputo, Davide and Ballarini, Dario and Dagvadorj, Galbadrakh and Sánchez Muñoz, Carlos and De Giorgi, Milena and Dominici, Lorenzo and West, Kenneth and Pfeiffer, Loren N. and Gigli, Giuseppe and Laussy, Fabrice P. and et al.},
   year={2017},
   month={Dec},
   pages={145–151}
}

@article{Dagvadorj15,
  title = {Nonequilibrium Phase Transition in a Two-Dimensional Driven Open Quantum System},
  author = {Dagvadorj, G. and Fellows, J. M. and Matyja\ifmmode \acute{s}\else \'{s}\fi{}kiewicz, S. and Marchetti, F. M. and Carusotto, I. and Szyma\ifmmode \acute{n}\else \'{n}\fi{}ska, M. H.},
  journal = {Phys. Rev. X},
  volume = {5},
  issue = {4},
  pages = {041028},
  numpages = {9},
  year = {2015},
  month = {Nov},
  publisher = {American Physical Society},
  doi = {10.1103/PhysRevX.5.041028},
  url = {https://link.aps.org/doi/10.1103/PhysRevX.5.041028}
}

@article{Donati16,
   title={Twist of generalized skyrmions and spin vortices in a polariton superfluid},
   volume={113},
   ISSN={1091-6490},
   url={http://dx.doi.org/10.1073/pnas.1610123114},
   doi={10.1073/pnas.1610123114},
   number={52},
   journal={Proceedings of the National Academy of Sciences},
   publisher={Proceedings of the National Academy of Sciences},
   author={Donati, Stefano and Dominici, Lorenzo and Dagvadorj, Galbadrakh and Ballarini, Dario and De Giorgi, Milena and Bramati, Alberto and Gigli, Giuseppe and Rubo, Yuri G. and Szymańska, Marzena Hanna and Sanvitto, Daniele},
   year={2016},
   month={Dec},
   pages={14926–14931}
}

@article{Dominici15,
   title={Vortex and half-vortex dynamics in a nonlinear spinor quantum fluid},
   volume={1},
   ISSN={2375-2548},
   url={http://dx.doi.org/10.1126/sciadv.1500807},
   doi={10.1126/sciadv.1500807},
   number={11},
   journal={Science Advances},
   publisher={American Association for the Advancement of Science (AAAS)},
   author={Dominici, Lorenzo and Dagvadorj, Galbadrakh and Fellows, Jonathan M. and Ballarini, Dario and De Giorgi, Milena and Marchetti, Francesca M. and Piccirillo, Bruno and Marrucci, Lorenzo and Bramati, Alberto and Gigli, Giuseppe and et al.},
   year={2015},
   month={Dec},
   pages={e1500807}
}

@article{Raimond01r,
  title = {Manipulating quantum entanglement with atoms and photons in a cavity},
  author = {Raimond, J. M. and Brune, M. and Haroche, S.},
  journal = {Rev. Mod. Phys.},
  volume = {73},
  issue = {3},
  pages = {565--582},
  numpages = {0},
  year = {2001},
  month = {Aug},
  publisher = {American Physical Society},
  doi = {10.1103/RevModPhys.73.565},
  url = {https://link.aps.org/doi/10.1103/RevModPhys.73.565}
}

@article{Walther06r,
	doi = {10.1088/0034-4885/69/5/r02},
	year = 2006,
	month = {apr},
	publisher = {{IOP} Publishing},
	volume = {69},
	number = {5},
	pages = {1325--1382},
	author = {Herbert Walther and Benjamin T H Varcoe and Berthold-Georg Englert and Thomas Becker},
	title = {Cavity quantum electrodynamics},
	journal = {Reports on Progress in Physics}
}

@article{Reiserer15r,
  title = {Cavity-based quantum networks with single atoms and optical photons},
  author = {Reiserer, Andreas and Rempe, Gerhard},
  journal = {Rev. Mod. Phys.},
  volume = {87},
  issue = {4},
  pages = {1379--1418},
  numpages = {40},
  year = {2015},
  month = {Dec},
  publisher = {American Physical Society},
  doi = {10.1103/RevModPhys.87.1379},
  url = {https://link.aps.org/doi/10.1103/RevModPhys.87.1379}
}

@article{Ng11,
  title = {{Exact real-time dynamics of quantum spin systems using the positive-P representation}},
  author = {Ng, Ray and S\o{}rensen, Erik S.},
  journal = {J. Phys. A},
  volume = {44},
  pages = {065305},
  year = {2011}
}

@article{Noh16,
	doi = {10.1088/0034-4885/80/1/016401},
	url = {https://doi.org/10.1088/0034-4885/80/1/016401},
	year = 2016,
	month = {nov},
	publisher = {{IOP} Publishing},
	volume = {80},
	number = {1},
	pages = {016401},
	author = {Changsuk Noh and Dimitris G Angelakis},
	title = {Quantum simulations and many-body physics with light},
	journal = {Reports on Progress in Physics}
}

@article{Lepert2011,
doi = {10.1088/1367-2630/13/11/113002},
url = {https://doi.org/10.1088/1367-2630/13/11/113002},
year = {2011},
month = {nov},
publisher = {IOP Publishing},
volume = {13},
number = {11},
pages = {113002},
author = {Lepert, G and Trupke, M and Hartmann, M J and Plenio, M B and Hinds, E A},
title = {Arrays of waveguide-coupled optical cavities that interact strongly with atoms},
journal = {New Journal of Physics}
}

@article{Hryniuk2024,
   title={{Tensor-network-based variational Monte Carlo approach to the non-equilibrium steady state of open quantum systems}},
   volume={8},
   ISSN={2521-327X},
   url={http://dx.doi.org/10.22331/q-2024-09-17-1475},
   DOI={10.22331/q-2024-09-17-1475},
   journal={Quantum},
   publisher={Verein zur Forderung des Open Access Publizierens in den Quantenwissenschaften},
   author={Hryniuk, Dawid A. and Szymańska, Marzena H.},
   year={2024},
   month={Sep}, pages={1475} }

@article{Zhu19,
doi = {10.1088/1367-2630/ab354d},
url = {https://doi.org/10.1088/1367-2630/ab354d},
year = {2019},
month = {aug},
publisher = {IOP Publishing},
volume = {21},
number = {8},
pages = {082001},
author = {Zhu, Bihui and Rey, Ana Maria and Schachenmayer, Johannes},
title = {A generalized phase space approach for solving quantum spin dynamics},
journal = {New Journal of Physics}
}

@article{Bond24,
    author = {Bond, Liam J. and Gerritsen, Bas and Minář, Jiří and Young, Jeremy T. and Schachenmayer, Johannes and Safavi-Naini, Arghavan},
    title = {{Open quantum dynamics with variational non-Gaussian states and the truncated Wigner approximation}},
    journal = {The Journal of Chemical Physics},
    volume = {161},
    number = {18},
    pages = {184113},
    year = {2024},
    month = {11},
    issn = {0021-9606},
    doi = {10.1063/5.0226268},
    url = {https://doi.org/10.1063/5.0226268}
}

@misc{Chuchurka23arxiv,
      title={{Quantum stochastic trajectories for particles and fields based on positive P-representation}}, 
      author={Stasis Chuchurka and Andrei Benediktovitch and Nina Rohringer},
      year={2023},
      eprint={2306.17474},
      archivePrefix={arXiv},
      primaryClass={quant-ph},
      url={https://arxiv.org/abs/2306.17474}, 
}

@article{Chuchurka24,
  title = {Stochastic modeling of x-ray superfluorescence},
  author = {Chuchurka, Stasis and Benediktovitch, Andrei and Kru\ifmmode \check{s}\else \v{s}\fi{}i\ifmmode \check{c}\else \v{c}\fi{}, {\ifmmode \check{S}\else \v{S}\fi{}} and Halavanau, Aliaksei and Rohringer, Nina},
  journal = {Phys. Rev. A},
  volume = {109},
  issue = {3},
  pages = {033725},
  numpages = {33},
  year = {2024},
  month = {Mar},
  publisher = {American Physical Society},
  doi = {10.1103/PhysRevA.109.033725},
  url = {https://link.aps.org/doi/10.1103/PhysRevA.109.033725}
}

@article{Chuchurka24b,
  title = {Stochastic modeling of superfluorescence in compact systems},
  author = {Chuchurka, Stasis and Sukharnikov, Vladislav and Benediktovitch, Andrei and Rohringer, Nina},
  journal = {Phys. Rev. A},
  volume = {110},
  issue = {5},
  pages = {053703},
  numpages = {24},
  year = {2024},
  month = {Nov},
  publisher = {American Physical Society},
  doi = {10.1103/PhysRevA.110.053703},
  url = {https://link.aps.org/doi/10.1103/PhysRevA.110.053703}
}

@misc{Leonau25arxiv,
      title={{Modelling many-body quantum dynamics with stochastic trajectories: a critical test on the Tavis-Cummings model}}, 
      author={A. Leonau and S. Chuchurka and V. Sukharnikov and A. Benediktovitch and N. Rohringer},
      year={2025},
      eprint={2511.08815},
      archivePrefix={arXiv},
      primaryClass={quant-ph},
      url={https://arxiv.org/abs/2511.08815}, 
}

@article{Hryniuk2026,
   title={Variational approach to open quantum systems with long-range competing interactions},
   volume={9},
   pages={45},
   ISSN={2399-3650},
   url={http://dx.doi.org/10.1038/s42005-025-02479-2},
   number={1},
   journal={Communications Physics},
   publisher={Springer Science and Business Media LLC},
   author={Hryniuk, Dawid A. and Szymańska, Marzena H.},
   year={2026},
   month=jan }

@misc{Hartmann2026arxiv,
      title={Truncated Wigner approximation for spins in continuous phase space}, 
      author={Jens Hartmann and Tom Schlegel and Viktoria Noel and Christopher D. Mink and Michael Fleischhauer},
      year={2026},
      eprint={2607.17295},
      archivePrefix={arXiv},
      primaryClass={quant-ph},
      url={https://arxiv.org/abs/2607.17295}, 
}

@article{Hosseinabadi25,
  title = {User-Friendly Truncated Wigner Approximation for Dissipative Spin Dynamics},
  author = {Hosseinabadi, Hossein and Chelpanova, Oksana and Marino, Jamir},
  journal = {PRX Quantum},
  volume = {6},
  issue = {3},
  pages = {030344},
  numpages = {23},
  year = {2025},
  month = {Sep},
  publisher = {American Physical Society},
  doi = {10.1103/1wwv-k7hg},
  url = {https://link.aps.org/doi/10.1103/1wwv-k7hg}
}

\end{document}